\documentclass[aps,twocolumn]{revtex4-1} 
\usepackage[dvipdfmx]{graphicx}
\usepackage{amsmath}
\usepackage{amssymb}
\usepackage{bm}
\usepackage{color}
\usepackage{ulem}

\usepackage{url}
\definecolor{red}{rgb}{1.0,0.0,0.0}

\def\bra#1{\mathinner{\langle{#1}|}}
\def\ket#1{\mathinner{|{#1}\rangle}}
\def\braket#1{\mathinner{\langle{#1}\rangle}}

\newcommand{\tfix}[1]{{\textcolor{black}{#1}}}

\newcommand{\ki}[1]{{\textcolor{black}{#1}}}

\begin{document}
%  \title{Charge dynamics of correlated electrons formulated by variational description \\
%with inclusion of composite fermions}
  \title{Charge dynamics of correlated electrons: Variational description\\ 
  with inclusion of composite fermions}
  \author{Kota Ido$^1$, Masatoshi Imada$^{2,3}$ and Takahiro Misawa$^1$}
  \affiliation{$^1$Institute for Solid State Physics, University of Tokyo, 
5-1-5 Kashiwanoha, Kashiwa, Chiba 277-8581, Japan\\
  $^2$Toyota Physical and Chemical Research Institute, Yokomichi, Nagakute, Aichi 480-1192, Japan\\
  $^3$Research Institute for Science and Engineering, Waseda University, 3-4-1, Okubo, Shinjuku, Tokyo 169-8555, Japan}

\begin{abstract}
 We propose a method to calculate the charge dynamical 
  structure factors for the ground states of correlated electron systems 
  based on the variational Monte Carlo method. 
  Our benchmarks for the one- and two-dimensional Hubbard models
  show that inclusion of composite-fermion excitations in the basis set
  greatly improves the accuracy, in reference to the exact
  charge dynamical structure factors for clusters.
  Together with examination for larger systems beyond tractable sizes by the exact diagonalization, our results indicate 
that the variational Monte Carlo method is a promising way for
  studies on the nature of charge dynamics in correlated materials 
  such as the copper oxide superconductors, if the composite-fermion excitations are properly included in the restricted Hilbert space 
  of intermediate states in the linear response theory. 
  \tfix{Our results are consistent with the particle-hole excitations inferred from the single-particle spectral function $A(\bm{k},\omega)$ in the literature.}
  We also discuss the importance of incorporating nonlocal composite fermion for more accurate description\tfix{. Future issues for further improvements are also discussed.} 
\end{abstract}

\maketitle
%%%%%%%%%%%%%%%%%%%%%%%%%%%%%%%%%%%%%%%%%%%%%%%%%%%%
%
\section{Introduction}
%
%%%%%%%%%%%%%%%%%%%%%%%%%%%%%%%%%%%%%%%%%%%%%%%%%%%%
Strongly correlated electron systems are 
a platform to search the emergent properties of nature 
such as the 
breakdown of single-particle descriptions manifested,
for instance, by non-Fermi liquid properties 
near Mott transitions\cite{Imada1998}, 
quantum spin liquids\cite{Balents2010,Zhou2017,Savary2017}, 
and unconventional high-$T_c$ superconductivity\cite{Keimer2015}.
To correctly understand the nature of the strongly correlated systems and 
control their physical properties, it is important to capture dominant part of elementary excitations  
emerging from correlations among electrons, which can be essentially different 
from what is expected from the noninteracting picture. 

To experimentally measure those excitations,
there exist several powerful techniques such as the 
neutron scattering, magnetic resonance, pump-probe optical measurement, 
angle-resolved photoemission spectroscopy (ARPES),
and resonant inelastic X-ray scattering (RIXS).
To directly compare the experimental results obtained 
by the above techniques with theoretical estimates,
it is necessary to develop accurate and efficient numerical methods 
to calculate  dynamical physical quantities. 

A seminal method for dynamical structure factors and dispersion of elementary bosonic excitations was formulated by Feynmann in his application to $^4$He, 
which succeeded in identifying rotons, in combination of neutron and X-ray measurement\cite{Feynman1954}.   
We follow basically the same spirit to calculate the dynamical structure factor
by constructing excited states from the variational ground states satisfying the variational principle,
now for strongly correlated fermionic systems. Thanks to the progress in computational power and methodology for dynamics since the Feyman's work, the formulation is much more sophisticated than the original form of Feynmann as we describe below.
 
As one of the methods,
analytical continuation of quantum Monte Carlo (QMC) data is often used 
for calculating dynamical structure factors.
In this approach, 
by using the QMC method, one estimates the 
real-frequency ($\omega$) spectrum 
from the imaginary-time ($\tau$) quantities 
such as the Green's function. 
It has been shown that this approach 
\tfix{provides numerically exact spectrum within statistical errors for quantum-many body systems}\cite{Silver1990,Jarrell1996,Sandvik1998}.
The analytical continuation is, however, an 
ill-posed problem and sensitive to noises in the calculated imaginary-time data.
In addition, the QMC method frequently suffers from the notorious 
negative-sign problem and its applicable range is limited.
Therefore, an alternative approach without 
the analytical continuation and the sign problem is desired.

The density matrix renormalization group (DMRG) method is 
a \tfix{numerical} method
for analyzing the quantum many-body systems 
in low dimensions without the sign problems\cite{White1992,Schollwock2005}.
By using the DMRG, the dynamical quantities
can be calculated efficiently in one dimension\cite{Hallberg1995,Ramasesha1997, Kuhner1999, Jeckelmann2000, Jeckelmann2002,Vidal2004,White2004,Daley2004}. 
One of the widely used approaches is the dynamical DMRG (DDMRG) method\cite{Jeckelmann2000, Jeckelmann2002},
where the real-frequency quantities are obtained from truncated density matrix for excitations.
Another approach is the time-dependent DMRG (tDMRG) method\cite{Vidal2004,White2004,Daley2004}.
In the tDMRG, 
one first obtains the real-time dependence of physical quantities from the direct real-time evolution.
The real frequency quantities can be calculated after its Fourier transformation.
However, the DMRG has difficulties in treating large systems in more than one dimension. 

The variational Monte Carlo (VMC) method 
is another powerful method free from the sign problem 
\cite{becca2017quantum}.
This method has been applied to analyze the ground states 
in a wide range of strongly correlated electron systems 
such as the two-dimensional Hubbard model and its extensions
\cite{Gros1987,Giamarchi1991,Sorella2001,Sorella2002, Yokoyama2013,Misawa2014a,Tocchio2013,Zhao2017,Ido2018}. 
Although the original applications of the VMC were limited to the ground states, 
a method of treating excited states and thermodynamic properties at nonzero temperatures were developed later 
by using approximated thermal pure states\cite{Takai2016} and 
transient states for nonequilibrium\cite{Carleo2012,Ido2015,Cevolani2015}.
In these approaches, based on the time-dependent variational principle\cite{Mclachlan1964,Haegeman2011},
the imaginary-time or the real-time evolutions of the variational
wave functions can be calculated.
In principle, 
as is the case for the tDMRG approach, 
the real-frequency spectrum can be obtained by using the Fourier transformations\cite{Ido2017}. 
To perform the accurate Fourier transformation, however,
the data of long real-time evolution is needed and the numerical cost becomes demanding. 
Therefore, a direct way of calculating the real-frequency properties 
within the VMC method is desired.

Recently, Li and Yang developed a 
VMC method to directly calculate the spin 
dynamical structure factor\cite{Li2010}.
The essence of their approach is to construct a relevant Hilbert 
subspace as the basis set by utilizing the Gutzwiller function applied to 
particle-hole excitations generated from the approximated ground state.
Within the restricted Hilbert subspace,
dynamical properties can be directly calculated.
Moreover, thanks to the Gutzwiller projection, 
their approach was shown to be capable of 
spinon dynamics in quantum spin chains\cite{Ferrari2018}
beyond simple particle-hole excitations. 
One might think that the Li-Yang method for quantum spin dynamics 
could be straightforwardly extended 
to charge dynamics such as the dynamical charge structure factor.

However, charge dynamics dominated by Mottness (barely itinerant electrons near the Mott insulator) together with interplay of spin fluctuations is 
a much more challenging subject. It involves competitions of 
charge, spin and superconducting order/fluctuations. 
If satisfactorily accurate methods are formulated, it has a wide
range of applications in itinerant electron systems.

In this paper, we propose a nontrivial and efficient way of extension to the Li-Yang method to allow computation of 
the charge dynamical structure factors
and test its accuracy by taking an example of the Hubbard model. 
Our benchmark calculations indicate that 
straightforward applications of the Li-Yang method 
with inclusion of only the simple particle-hole excitations
do not 
reproduce satisfactory
charge dynamical structure factor even with Gutzwiller projection.
By contrast, we show that the inclusion of the composite fermions into the basis set of the restricted subspace offers much better description of 
the charge dynamics. It is favorably compared with
the exact diagonalization and the DMRG results.

The paper is organized as follows. 
In Sec. \ref{sec:method}, we first briefly review the essence of
the Li-Yang method for dynamical quantities, which is based on the VMC method. 
Then we present our extension of composite-fermion approach 
to calculate the charge dynamics.
Section \ref{sec:benchmarks} presents our benchmarks for the 
charge dynamical structure factor for the Hubbard model. 
It is shown that introduction of the composite fermions drastically improves
the charge dynamics in the Hubbard model.
In the one-dimensional Hubbard model, we find that our method quantitatively reproduces
the results of the exact diagonalization.
Our results for larger sized systems in one dimension are also  
consistent with the previous tDMRG result\cite{Pereira2012}.
Application to the two-dimensional Hubbard model shows 
\tfix{that our method also largely improves accuracy of 
the charge dynamics compared with the bare-fermion approach 
}
\tfix{qualitatively} representing \tfix{exact} charge dynamics 
of the undoped and doped Mott insulators 
including the incoherent part of dynamics.
We, however, point out the possible further improvement \tfix{for the quantitative accuracy in two dimensional systems} by incorporating additional composite fermions not taken into account in this paper.
Finally, in Sec. \ref{sec:summary} we summarize our paper and discuss future directions.

%%%%%%%%%%%%%%%%%%%%%%%%%%%%%%%%%%%%%%%%%%%%%%%%%%%
%
\section{Method}\label{sec:method}
%
%%%%%%%%%%%%%%%%%%%%%%%%%%%%%%%%%%%%%%%%%%%%%%%%%%%%
\subsection{Variational approach for excited states}
In this subsection, we review a variational approach 
to calculate the dynamical spin structure factor $S(\bm{q},\omega)$.
This method was introduced by Li and Yang to obtain $S(\bm{q},\omega)$ in the $t$-$J$ model\cite{Li2010}, 
and it has been recently applied to 
the one-~\cite{Ferrari2018} and two-dimensional Heisenberg model~\cite{DallaPiazza2015,Ferrari2018a}.
The idea of this method is to restrict the Hilbert space 
to a set of $\ket{\bm{q},n}$, which has momentum $\bm{q}$ with an 
index $n$ which specifies type of excitations.
In general, the basis set $\{ \ket{\bm{q},n} \}$ on the restricted 
Hilbert subspace is nonorthogonal and thus the generalized eigenvalue 
problem within this subspace can be written as
\begin{align}
  \sum_{m} H^{\bm{q}}_{n,m} v^{\bm{q},l}_{m} = E_{\bm{q},l}^{\rm var} \sum_{m} O^{\bm{q}}_{n,m} v^{\bm{q},l}_{m}, \label{gep}
\end{align}
Here, the matrix element of Hamiltonian $\mathcal{H}$ and overlap matrices on the subspace are represented as 
\begin{align}
  H^{\bm{q}}_{n,m} &= \frac{ \braket{ \bm{q},n | \mathcal{H} | \bm{q},m}}{\braket{\psi | \psi}}, \label{rh} \\
  O^{\bm{q}}_{n,m} &= \frac{ \braket{ \bm{q},n | \bm{q},m}}{\braket{\psi | \psi}}. \label{ro}
\end{align}
By solving this generalized eigenvalue problem defined in Eq.~(\ref{gep}), we can obtain the $l$-th 
eigenvalue $E^{\bm{q},l}_{\rm var}$ and the coefficients of its eigenvector $v^{\bm{q},l}_{m}$.
In other words, the excited state $\ket{\psi^{\rm var}_{\bm{q},l}}$ within the 
subspace is expressed in the following equation:
\begin{align}
  \ket{\psi^{\rm var}_{\bm{q},l}} = \sum_{n} v^{\bm{q},l}_{n} \ket{\bm{q},n}.
\end{align}

Based on this approach, the dynamical spin structure factor of $\alpha=x,y,z$ component of spin, $S^{\alpha}(\bm{q},\omega)$ can be computed.
For example, $S^z(\bm{q},\omega)$ is described as 
\begin{align}
  S^z(\bm{q},\omega) &= -\frac{1}{\pi} {\rm Im} \braket{\overline{\psi} | S^z_{-\bm{q}}\frac{1}{\omega-\mathcal{H}+i\eta} S^z_{\bm{q}}| \overline{\psi}}, \label{sz} \\
  S^z_{\bm{q}} &=  \frac{1}{\sqrt{N_s}} \sum_{j} \exp (-i\bm{q} \cdot \bm{r}_j) S^z_{j},
\end{align}
where $\ket{\overline{\psi}}$ represents 
the normalized ground-state wavefunction and its energy, respectively. 
Hereafter, we assume that the total momentum of the ground state is zero. 
We note that $\eta$ is a phenomenological smearing factor. 
By inserting the complete set 
$\sum_l  \ket{\psi^{\rm var}_{\bm{q},l}}  \bra{\psi^{\rm var}_{\bm{q},l}} $
of momentum $\bm{q}$ in the restricted Hilbert space into Eq. (\ref{sz}), 
we obtain
\begin{align}
  S^z(\bm{q},\omega) &\approx& \frac{1}{\pi} \sum_{l} \left| \braket{\psi^{\rm var}_{\bm{q},l} | S^z_{\bm{q}}| \overline{\psi}} \right|^2 \frac{\eta}{(\omega-E_{\bm{q},l}^{\rm var})^2+\eta^2}. \nonumber \label{sz2} \\
\end{align}

To obtain accurate $S(\bm{q},\omega)$, choice of the basis set, namely how to pick up the restricted Hilbert space, is important.
A simplest choice is
\begin{align}
  \ket{\bm{q},0} = S^z_{\bm{q}}\ket{\psi} = \frac{1}{\sqrt{N_s}} \sum_{j,\sigma} \exp \left[ -i\bm{q}\cdot \bm{r}_j \right] \sigma n_{j\sigma} \ket{\psi}, \label{sma}
\end{align}
where $n_{i\sigma}=c^{\dagger}_{i\sigma} c_{i\sigma}$ and $c^{\dagger}_{i\sigma}$ ($c_{i\sigma}$) is a creation operator for an electron with spin $\sigma$ at position $\bm{r}_i$.
Solution obtained only from this restricted Hilbert space is called the single-mode approximation. 
In that case, one can easily compute the pole position $E_{\bm{q},0}^{\rm var}=H^{\bm{q}}_{0,0}/O^{\bm{q}}_{0,0}$ 
because the dimension of the subspace at $\bm{q}$ is only one. 
However, this approach can capture only an isolated dispersion, 
and is not able to represent 
continuum spectra if it exists as is expected in correlated systems. 
To overcome this limitation, 
the Gutzwiller wave function with 
particle-hole excitations is employed
as the basis set of the restricted Hilbert space
in the previous studies~\cite{Li2010,DallaPiazza2015,Ferrari2018}.
This extension is given as
\begin{align}
  \ket{\bm{q},R} = \mathcal{P} \frac{1}{\sqrt{N_s}} \sum_{j,\sigma} \exp \left[ -i\bm{q}\cdot \bm{r}_j \right] \sigma c^{\dagger}_{j+R,\sigma} c_{j\sigma} \ket{\phi}, \label{spinon}
\end{align}
where $\mathcal{P}$ represents the Gutzwiller factor to exclude the doubly occupied sites
and $\ket{\phi}$ is a mean-field solution for the ground state~\cite{Ferrari2018}.
\tfix{$R = 1,2,\cdots N_s$ are the indices of the lattice coordinates.}
It was shown that the restricted basis set (\ref{spinon}) can 
well describe the spinon continuum of $S(\bm{q},\omega)$ in the one-dimensional 
Heisenberg model and in its extension \tfix{compared with the exact results}\cite{Ferrari2018}.

Advantages of this method are summarized as follows:
\begin{enumerate}
\item No negative sign problems.
\item Analytical continuation is {\it not} necessary.
\item Excited states can be explicitly constructed from the ground-state wave function.
\end{enumerate}
\tfix{The third advantage can be rephrased as mechanisms and intuitive understanding of dynamical properties are figured out and extracted concretely from explicit forms of the wavefunctions, by comparing the two cases with and without each specific part of the wave functions. It clarifies how taking proper component of the excited states into account is important to reach accurate dynamical quantities.}
As we show in the rest of this paper,
incorporating the composite-fermion excitations
into the restricted Hilbert space is essential for accurate description of charge dynamical structure factors.

%%%%
\subsection{Construction of excited states for charge dynamics}
%%%%
Following the idea in the previous studies, we propose a way to reasonably approximate the charge 
structure factor $N(\bm{q},\omega)$ 
in the strongly correlated itinerant electron systems
such as the Hubbard model.
The definition of $N(\bm{q},\omega)$ is
\begin{align}
  N(\bm{q},\omega) &= -\frac{1}{\pi} {\rm Im} \braket{\overline{\psi}| n_{-\bm{q}} \frac{1}{\omega-\mathcal{H}+i\eta}  n_{\bm{q}} | \overline{\psi}} \\
  &= \frac{1}{\pi} \sum_l \left| \braket{\psi_{\bm{q},l} | n_{\bm{q}} | 
  \overline{\psi} } \right|^2 \frac{\eta}{\left( \omega -E_{{\bm{q},l}}\right)^2 + \eta^2}
\end{align}
where
\begin{align}
  n_{\bm{q}} &= \frac{1}{N_s} \sum_j \exp \left[ -i\bm{q}\cdot \bm{r}_j \right] \left( n_{j\uparrow} + n_{j\downarrow} \right).
\end{align}
In order to calculate $N(\bm{q},\omega)$ in the VMC method, we approximate $N(\bm{q},\omega)$ in the same manner as Eq. (\ref{sz2}), namely
\begin{align}
  N(\bm{q},\omega) &\approx \frac{1}{\pi} \sum_l \left| \braket{\psi^{\rm var}_{\bm{q},l} | n_{\bm{q}} | \overline{\psi} } \right|^2 \frac{\eta}{\left( \omega -E^{\rm var}_{{\bm{q},l}}\right)^2 + \eta^2}. 
\end{align}
We note that, in this scheme using the restricted Hilbert space, the sum rule is satisfied. 
Therefore, the integral of the approximated 
$N(\bm{q},\omega)$ over $\omega$ is reduced to the static (equal-time) charge structure factor, namely
\begin{align}
  \int d\omega N(\bm{q},\omega) &\approx \sum_l \left| \braket{\psi^{\rm var}_{\bm{q},l} | n_{\bm{q}} | \overline{\psi} } \right|^2 \nonumber \\
  &= \braket{\overline{\psi} | n_{-\bm{q}}  n_{\bm{q}} | \overline{\psi} } = N(\bm{q}). \nonumber \\
\end{align}

Next, we propose a way to restrict the basis set for $N(\bm{q},\omega)$.
A naive candidate in analogy with $S(\bm{q},\omega)$ would be 
\begin{align}
  \ket{\bm{q},R} = \frac{1}{\sqrt{N_s}} \sum_{j,\sigma} \exp \left[ -i\bm{q}\cdot \bm{r}_j \right] c^{\dagger}_{j+R,\sigma} c_{j\sigma} \ket{\psi}. \label{pair}
\end{align}
This approximation is the single-mode approximation of the 
density operator, $n_{\bm{q}}\ket{\psi}$, containing the 
long-range particle-hole excitation independent of spin degrees of freedom $\sigma$.
Throughout this paper, the VMC approach where Eq. (\ref{pair}) is employed as the basis set of the restricted Hilbert space is called the bare-fermion (BF) approach.
At first glance, it looks possible to represent proper particle-hole 
continuum in $N(\bm{q},\omega)$ as the case of $S^z(\bm{q},\omega)$.
However, as we show later, 
this simple choice does not work well for describing the charge 
dynamics in the Mott insulating states. 
This is because holons and doublons represent 
essential charge states in the Mott insulating state of the Hubbard model,
generating upper and lower Hubbard bands, which is not represented directly by
bare electron operators $c^\dagger$ and $c$.
The charge spectrum in the strong 
coupling limit 
is dominated 
by kinetics of excited holons and doublons in Mott-Hubbard bands.
To take this kinetics into account, 
we need to consider electron dynamics 
which depends on the occupation number of the sites 
in the process before and after the hopping. 
The importance of the separation of the electron kinetics 
has been already pointed out in the strong coupling approach 
based on the $S$-matrix expansion~\cite{Gros1987,MacDonald1990}.
It is also related to the electron fractionalization, which is exactly proven in the strong coupling limit~\cite{Zhu2013,Imada2019}.

Thus, we need to distinguish these electron fractionalization effects beyond the bare electron to 
capture correct physical description of charge dynamics of correlated electrons.
We therefore introduce composite fermion operators $z_{i\sigma \alpha}$, defined as
\begin{align}
  z_{i\sigma+} &= c_{i\sigma} n_{i\overline{\sigma}}, \label{z_p} \\
  z_{i\sigma-} &= c_{i\sigma} \left( 1-n_{i\overline{\sigma}} \right),
\label{z_m}
\end{align}
in addition to the bare electron operator
\begin{align}
  z_{i\sigma0} &= c_{i\sigma},
\end{align}  
where $\overline{\sigma}$ denotes the opposite spin of $\sigma$. 
The composite fermions $z_{i\sigma \pm}$ are known as 
the Hubbard operators\cite{Hubbard1965,Mancini2004}.
Since $z_{i\sigma +(-)}$ creates a 
spinon (holon) and $z_{i\sigma +(-)}^{\dagger}$ creates a doublon (spinon), 
they allow to differentiate dynamics of these particles. 
The importance of composite fermions to 
describe the nature of electronic structure in 
correlated materials has also been discussed in the structure of the single-particle spectral function\cite{Yamaji2011,Yamaji2011a,Zhu2013,Sakai2016a,Sakai2016,Imada2019}.
We further note that $z_{i\sigma0}$ can be expressed as 
$z_{i\sigma0}= z_{i\sigma+} + z_{i\sigma -}$ and we need only two of these
three because they are linearly dependent.

By using these operators, we here propose another 
candidate of the basis set as follows: 
\begin{align}
  \ket{\bm{q},R,\alpha,\beta} = \frac{1}{\sqrt{N_s}} \sum_{j,\sigma} \exp \left[ -i\bm{q}\cdot \bm{r}_j \right] z^{\dagger}_{j+R,\sigma,\alpha} z_{j \sigma \beta} \ket{\psi}. \nonumber \\ \label{new}
\end{align}
We call this VMC approach combined with the concept of composite-fermions as the composite-fermion (CF) approach.
We evaluate the Hamiltonian matrix within this restricted Hilbert space and the overlap matrix on the target subspace, defined in Eqs. (\ref{rh}) and (\ref{ro}), 
by using the reweighting technique for efficient Monte Carlo samplings\cite{Li2010}. 
See also Appendix \ref{appendixA} for the detail of the reweighting technique.

%%%%%%%%%%%%%%%%%%%%%%%%%%%%%%%%%%%%%%%%%%%%%%%%%%%
%
\section{Results}\label{sec:benchmarks}
%
%%%%%%%%%%%%%%%%%%%%%%%%%%%%%%%%%%%%%%%%%%%%%%%%%%%%
%%%%%%%%%%%%%%%%%%%%%%%%%%%%%%%%%%%%%%%%%%%%%%%%%%%
%
\subsection{Model and setting}
%
%%%%%%%%%%%%%%%%%%%%%%%%%%%%%%%%%%%%%%%%%%%%%%%%%%%%
As benchmarks, we performed simulations of the 
dynamical structure factors in the Hubbard model.
The Hubbard model is defined as
\begin{align}
  \mathcal{H} = -t\sum_{\braket{i,j},\sigma} c^\dagger_{i\sigma}c_{j\sigma} + U\sum_{i} n_{i\uparrow} n_{i\downarrow}, \label{hubbard}
\end{align}
where $t$ and $U$ are the hopping integral and the on-site Coulomb repulsion, respectively.
In the following results, we set the energy unit to the hopping integral ($t=1$).
We employed the smearing factor $\eta=0.2$ for demonstration of $N(\bm{q},\omega)$. 
\tfix{We note that this value is smaller than the effective exchange interaction induced by the strong onsite interaction.
See also Appendix B, where $\eta$-dependence of $N(\bm{q},\omega)$ is shown.}
We note that there exists a trivial strong 
peak at $\omega=0$ for $\bm{q}=\bm{0}$, 
whose integrated spectral weight is nothing but the square of the total number of particles, i.e, 
$N(\bm{q}= \bm{0},\omega)=N_{\rm e}^2 \delta(\omega)$ for $\eta \rightarrow 0$.
To eliminate this trivial 
peak and enhance visibility 
of the spectrum for other $\bm{q}$, 
we impose $N(\bm{q}=\bm{0},\omega)=0$.

The trial wavefunction for the ground state 
we employed is a pair product wave function\cite{Tahara2008a} 
with Gutzwiller-Jastrow\cite{Gutzwiller1963,Jastrow1955} correlation factors
\begin{align}
&\ket{\psi}=\mathcal{P_G}\mathcal{P_J}\ket{\phi},\\
  &\ket{\phi}=\left( \sum_{i,j} f_{ij} c^\dagger_{i\uparrow} c^\dagger_{j\downarrow} \right)^{N_{\rm e}/2} \ket{0},\\
&\mathcal{P_G}=\exp \left( g\sum_i n_{i\uparrow} n_{i\downarrow}\right),\\
&\mathcal{P_J}=\exp \left( \sum_{i,j} v_{ij} n_{i} n_{j} \right).
\end{align}
The variational parameters $f_{ij}$, $g$ and $v_{ij}$ 
are simultaneously optimized by using the stochastic reconfiguration method\cite{Sorella2001}.
We imposed the translational symmetry to the variational parameters 
for the one-dimensional case in order to reduce the numerical costs.
This assumption works well for one-dimensional systems 
because the ground state should not spontaneously break the symmetry 
even in the thermodynamic limit. 
Note that our trial wavefunction can represent the Mott insulator 
owing to the doublon-holon binding correlations in the Jastrow factor\cite{Capello2005}.
On the other hand, for the two dimensional case, 
we impose $2 \times 2$ sublattice structure on $f_{ij}$ to 
allow description of the antiferromagnetic states.

%%%%%%%%%%%%%%%%%%%%%%%%%%%%%%%%%%%%%%%%%%%%%%%%%%%
%
\subsection{One dimensional case}
%
%%%%%%%%%%%%%%%%%%%%%%%%%%%%%%%%%%%%%%%%%%%%%%%%%%%%
\begin{figure}[htbp]
  \begin{center}
    \includegraphics[width=80mm]{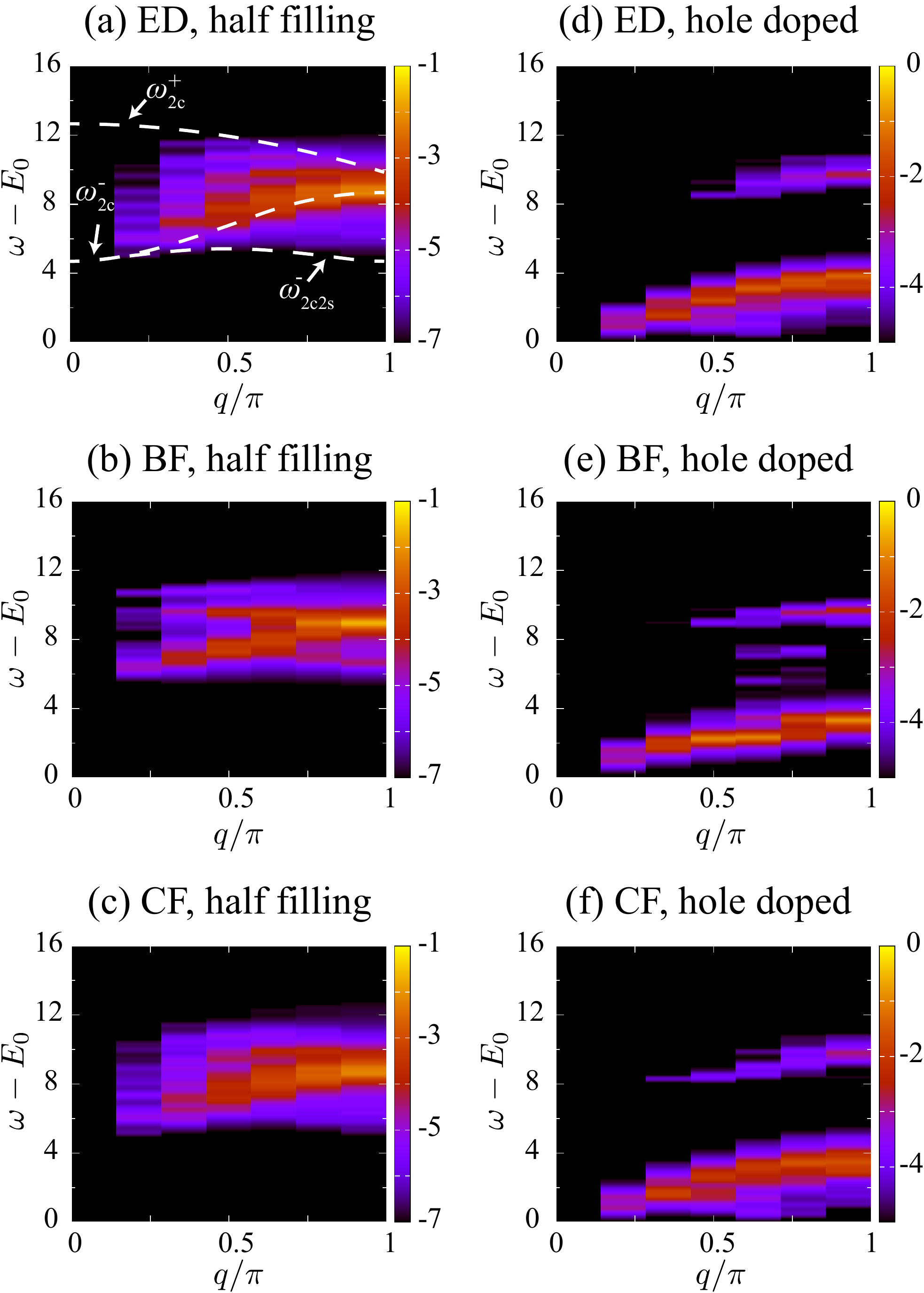}
  \end{center}
  \caption{(Color online) 
  Contour plot of the charge dynamical structure factor $\log N(\bm{q},\omega)$ obtained by the VMC method in the two different approaches for the one-dimensional Hubbard model with $L=14$, $U/t=8$.
  The ED results are also shown for comparison.
  \tfix{$E_0$ means the ground-state energy, i.e. $E_0 = \braket{\overline{\psi}| \mathcal{H}| \overline{\psi}}$.}
  To enhance the visibility for difference between the results obtained by using two different variational Ans\"{a}tze, the results are plotted in the logarithmic scale.
  The results for $N_{\rm e}=14$ and $N_{\rm e}=10$ are shown in panels (a-c) and (d-f), respectively.
  Panels (a) and (d) show the results obtained by using the ED method.
  In panels (b) and (e), we show the result obtained by the BF approach. 
  The results by the CF approach are shown in panels (c) and (f).
  White dashed curves in panel (a) are dispersions characterizing the Bethe Ansatz solution: 
%We draw some special lines of the spectrum characterized by the Bethe Ansatz solution at half filling as . 
  $\omega_{2c}^{+(-)}$ represents the upper (lower) edge energy of the two-holon [or two-doublon] continuum.
  $\omega_{2c2s}^{-}$ represents the lower edge energy of the two-holon-two-spinon [or two-doublon-two-spinon] continuum.
  Notations of these modes are the same as those in the previous study \cite{Pereira2012}.
  }
  \label{comparison}
\end{figure}

In this subsection, we show benchmarks of the charge dynamical 
structure factor in the one dimensional Hubbard model.
We mainly study the system with linear size $L=14$ under the periodic boundary condition
to directly compare with the exact diagonalization (ED) results.
In the last part of this subsection, we show the result for $L=50$ as well to
examine the applicable range of our method.
At half filling, we employ a staggered particle-hole transformation, 
$c_{i\uparrow} \rightarrow c^\dagger_{i\uparrow}$ 
and $c_{i\downarrow} \rightarrow (-1)^ic_{i\downarrow}^{\dagger}$, 
in order to improve the accuracy of the ground state\cite{Shiba1972}. 
We confirmed that the results do not change within the statistical error 
even if the transformation is not taken, although the statistical error is much 
higher.

\begin{figure}[htbp]
  \begin{center}
   \includegraphics[width=80mm]{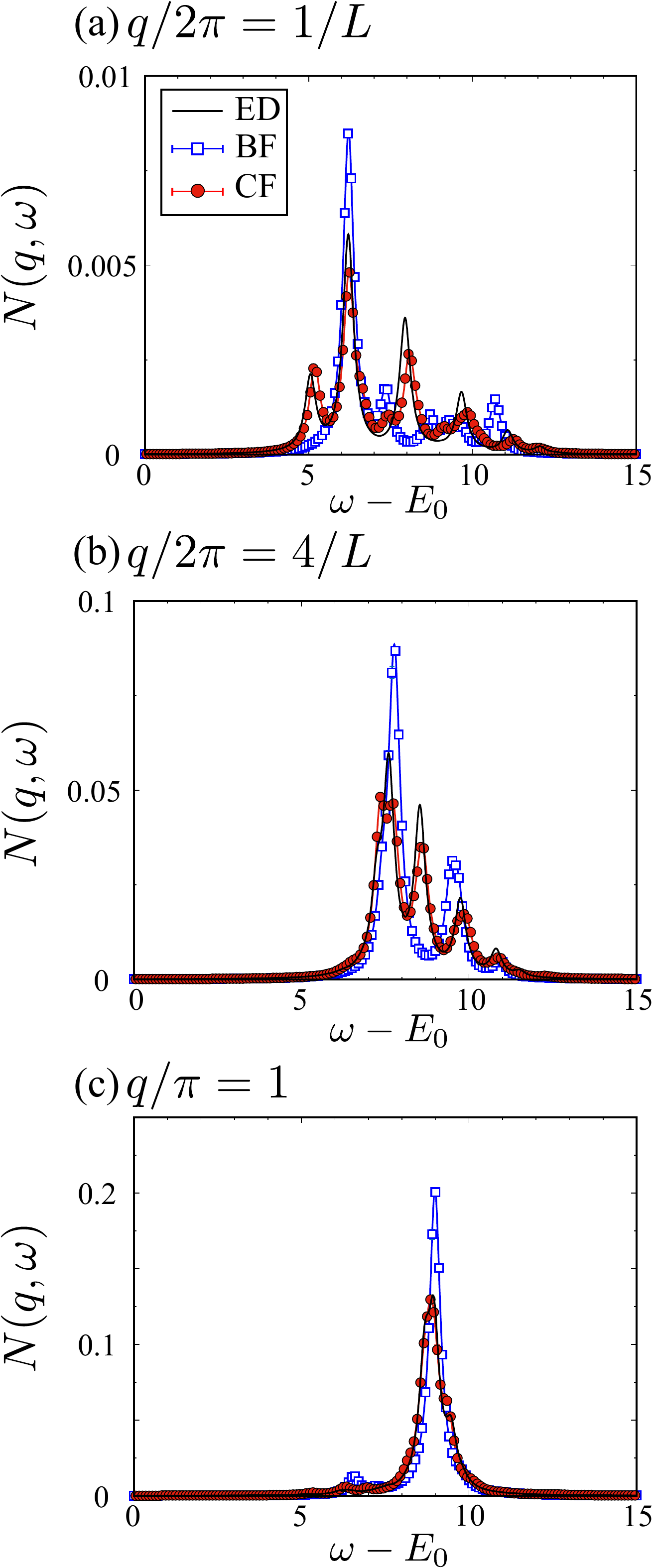}
  \end{center}
  \caption{
    (Color online) Charge dynamical structure factor in the one-dimensional Hubbard model for $L=14$ and $U/t=8$ at half filling for several choices of momentum $q$.
  Black solid lines are the results obtained by using the ED method.
  Open squares and solid circles represent the VMC results obtained by using the basis set in Eq. (\ref{pair}) and Eq. (\ref{new}), respectively. 
  }
  \label{qdep_chain}
\end{figure}

First, to clarify whether composite fermions play an essential role to describe exact charge dynamics, we show the color contour plot of $q$ and $\omega$ dependence of $N(q,\omega)$ 
obtained by two different variational Ans\"{a}tze and compare with the ED results for $U/t=8$ in Fig. \ref{comparison}.
At half filling, we draw three special modes obtained from the Bethe Ansatz 
equation\cite{Pereira2012}, $\omega_{2c}^{\pm}$ and $\omega_{2c2s}^{-}$, in panel (a) by white dashed curves.
In panel (a), there are two broad continuum at $U/t=8$ in the exact result.
One is the strong two-holon (or two-doublon) continuum between $\omega_{2c}^{+}$ and $\omega_{2c}^{-}$, and the other is the
weak two-holon-two-spinon (or two-doublon-two-spinon) continuum between $\omega_{2c}^{-}$ and $\omega_{2c2s}^{-}$ for $q/\pi \geq 0.5$.
Panel (b) shows the spectrum at half filling by using 
only the BF excitation in Eq. (\ref{pair}).
We found that the BF 
approach is able to describe two-holon-two-spinon continuum. 
This is reasonable because this continuum directly connects to 
particle-hole continuum in the limit of 
non-interacting system, which was mentioned 
in the previous DMRG calculation \cite{Pereira2012}.
However, the weight contributed from this continuum is larger than the exact one 
as we will detail in Fig.~\ref{qdep_chain}.
In addition, this approach predicts a strong lower edge $\omega_{2c}^{-}$ below the two-holon continuum. However, the ED results indicate that this region is 
dominated by a broad continuum especially for $q/\pi \leq 0.5$ and no sharp edge is observed.
The failure of capturing the broad two-holon continuum suggests that the BF approach does not describe the doublon-holon 
recombination process contributed from an excited doublon-holon pair.
On the other hand, as we see in panel (c), 
the CF approach well reproduces the ED result. 
\tfix{The main peaks dispersed from $\omega-E_0\sim 4$ at $q=0$ to 9 at $q=\pi$ is also consistent with 
the particle-hole excitation inferred from the spectral function $A(k,\omega)$ obtained in Refs. \cite{Kim1996, Kim1997, Charlebois2019}, 
where the splitting of $\omega_{2c}^{\pm}$ is identified as the excitation from (to) the spinon and holon branches.}

Next, we show the results for the hole-doped case in Fig.~\ref{comparison} (d)-(f).
Carrier doping eventually destroys the Mott gap, while the upper and 
lower Hubbard bands remain separated by the gap if the doping level is low. 
At low doping, the ground state becomes a correlated metal, where low-energy
excitation emerges near the Fermi level, but stays largely incoherent with broad feature particularly at large $q/\pi>0.5$. 
In addition, although the incoherent excitation 
exists around $\omega-E_0=U/t$ as 
a remnant of the Mott gap at half filling, its weight is greatly reduced
from the insulating case at higher doping level, which is transferred to the low-energy part.
The spectrum for finite doping calculated by using the 
BF approach is also shown in the panel (e).
It underestimates the broadness at large $q$.
The reason would be that the hole carrier doping creates 
not only holons but also spinons due to the motion of the induced holons.
Since the holon and spinon move separately because of the spin-charge separation, we need to distinguish dynamics of these particles,
which is beyond the representability of the single electron dynamics.
On the other hand, the composite fermion scheme improves this broadness.
The present results both for half filled and the hole doped cases indicate that the inclusion of the
composite fermions are required to describe correct charge dynamics of correlated electrons.

\begin{figure}[htbp]
  \begin{center}
   \includegraphics[width=80mm]{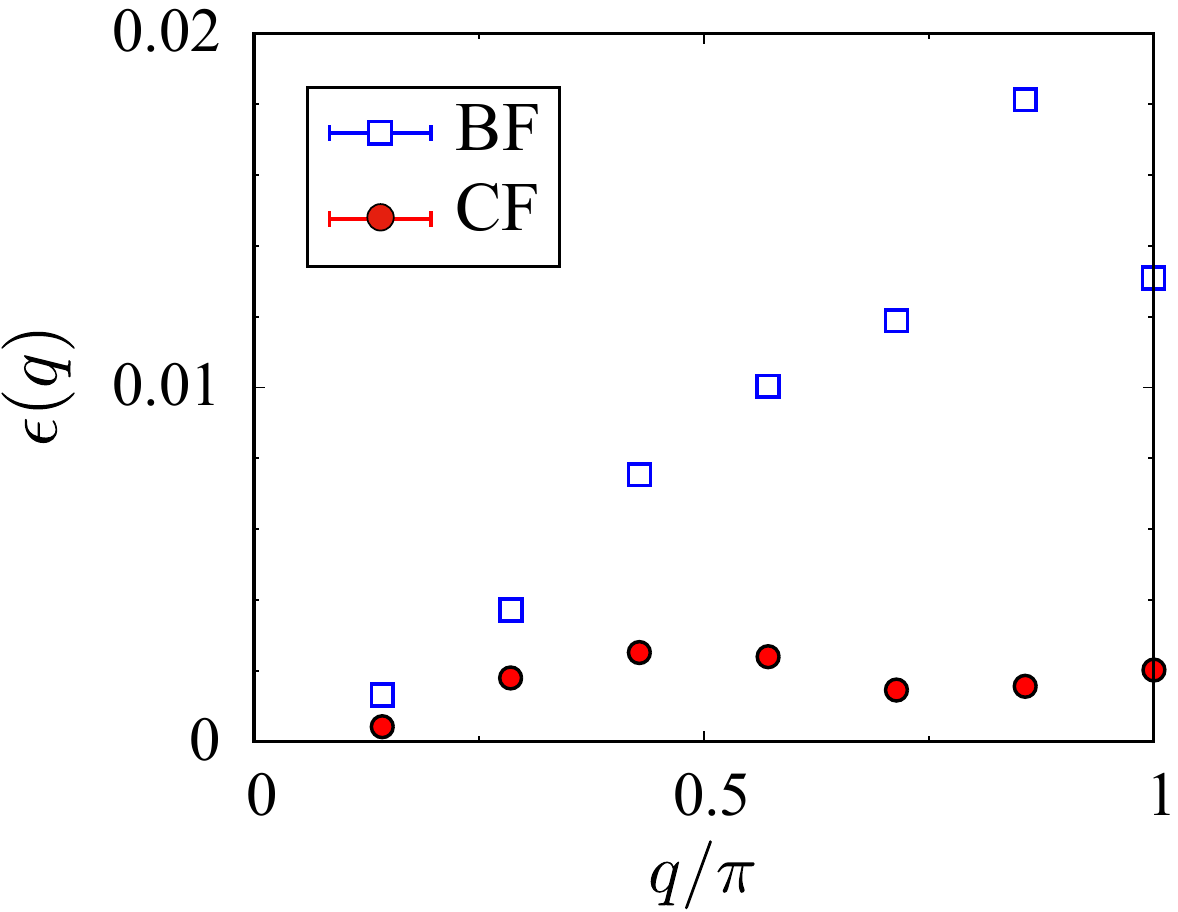}
  \end{center}
  \caption{
    (Color online) Momentum-dependence of the absolute error of the charge dynamical structure factor in the one-dimensional Hubbard model for $L=14$ and $U/t=8$ at half filling.
  Open squares (solid circles) represents the results in the BF (CF) scheme. 
  }
  \label{qdep_error}
\end{figure}

The superiority of the CF description is more clearly shown in the following analysis.
In Fig.~\ref{qdep_chain} we show $\omega$-dependence of 
the charge structure factors for several momenta at half filling. 
At a small momentum ($q/\ki{2}\pi=1/L$) [shown in Fig.~\ref{qdep_chain}(a)],
the spectrum obtained by the CF excitations
well reproduces the ED result.
The agreement of $N(q,\omega)$ at the small momentum indicates that our CF approach has possibility to describe the correct optical conductivity $\sigma(\omega)$ as well
because $\sigma(\omega)$ is tightly connected to the charge dynamical structure 
factor at $q \rightarrow 0$, namely $\sigma(\omega) = \omega \lim_{q\rightarrow 0} [N(q,\omega)/q^2]$~\cite{Stephan1996,Kim2004,Benthien2007}.
It is remarkable that the spinon excitations as a consequence of 
many-body effects in one dimension can essentially 
be captured by just two modes in the composite fermion scheme if the Gutzwiller and Jastrow factors are considered.
The spectrum with the BF excitations, however,
only reproduces the peak around $\omega -E_0 \sim 6$ and fails in representing other peaks.
This contrast between the BF and CF is similar for larger $q$ 
[shown in Figs.~\ref{qdep_chain}(b),(c)]:
The spectrum by BF excitations reproduces only
one peak, which corresponds to the lower-edge of the two-holon continuum $\omega_{2c}^{-}$. 
Furthermore, the amplitudes of the peak obtained by the BF excitations are larger than the exact results. 

To analyze the difference between two approaches more quantitatively, Fig. \ref{qdep_error} shows the absolute error of the charge dynamical structure factor for $U/t=8$ which is defined by
\begin{align}
  \epsilon(\bm{q}) = \frac{1}{N_{\rm \omega}} \sum_{n}^{N_{\rm \omega}} \left| N_{\rm ED} (\bm{q},\omega_n) - N_{\rm VMC} (\bm{q},\omega_n) \right| . \label{error_q}
\end{align}
Here $N_{\rm ED(VMC)}$ is the charge dynamical structure factor obtained by using the exact diagonalization (VMC) method, and $N_{\rm \omega}$ is the number of gird points on the $\omega$-line.
We can confirm 
that $\epsilon(\bm{q})$ by the CF approach is improved compared with that by the BF one.

\begin{figure}[tbp]
  \begin{center}
   \includegraphics[width=80mm]{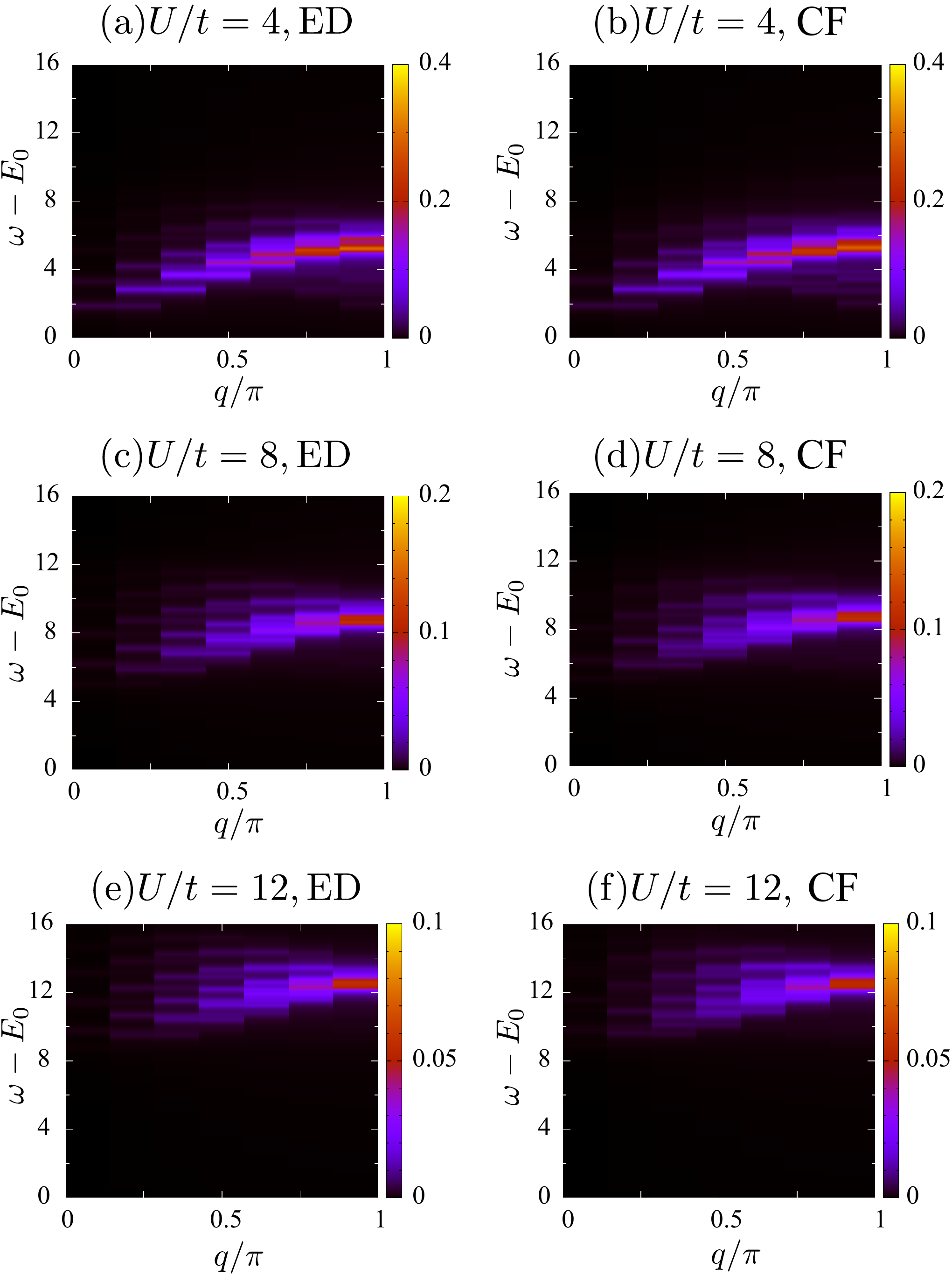}
  \end{center}
  \caption{(Color online) Contour plot of the charge dynamical structure factor in the one-dimensional Hubbard model for $L=14$ at half filling.
  The numerical method we used and the strength of the on-site interaction $U/t$ are described in the title of each panel.
  ``ED" and ``CF" represent the exact diagonalization method and the CF approach in the VMC method, respectively. 
  }
  \label{chain_hf}
\end{figure}

\begin{figure}[htbp]
  \begin{center}
   \includegraphics[width=80mm]{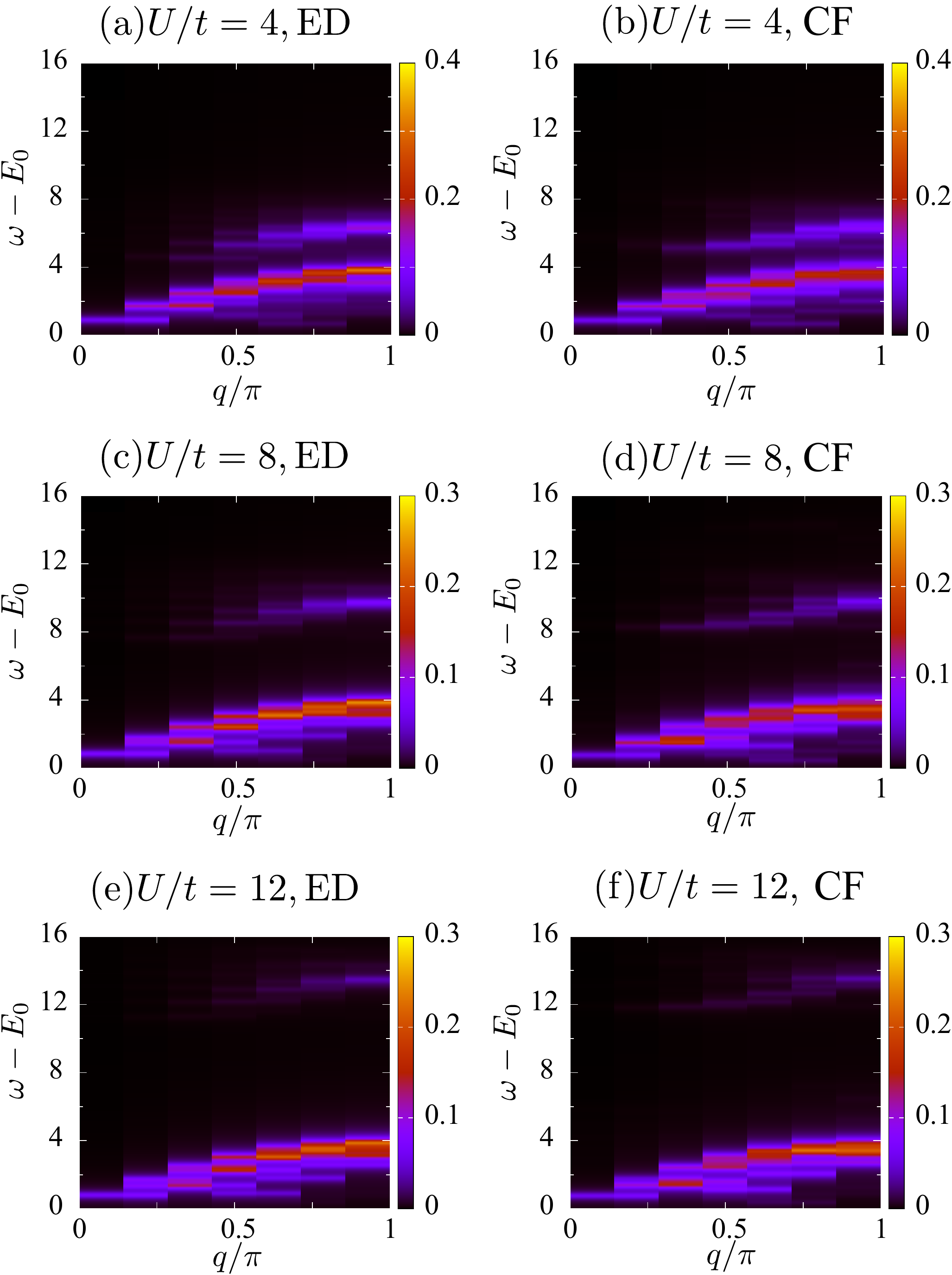}
  \end{center}
  \caption{(Color online) Contour plot of the charge dynamical structure factor in the hole-doped Hubbard chain for $L=14$ and $N_{\rm e}=10$.
  The numerical method we used and the strength of the on-site interaction $U/t$ are described in the title of each panel.
  ``ED" and ``CF" denote the exact diagonalization method and the CF approach in the VMC method, respectively. 
  }
  \label{chain_dope}
\end{figure}

Figures \ref{chain_hf} and \ref{chain_dope} show the 
interaction dependence of $N(\bm{q},\omega)$ in the Hubbard chain 
at half filling and the hole-doped case, respectively.
In these figures, we only show $N(\bm{q},\omega)$  by
the ED and the CF approach because
the accuracy of the BF approach
turned out to be poor as we mentioned above.
We see that in all the regions from weak to strong coupling, 
our VMC approach well reproduces the Mott gap scaled by $U/t$, which is manifested in the 
global shift of the structure in the $U$ dependence of $N(\bm{q},\omega)$. 
For hole doped case, we find that 
the low-energy excitation is not sensitive to the strength of $U$, and the CF approach captures the $U$ dependence well, including the broad incoherent feature, suggesting the correct description of the holon dynamics. 
The relative weight of the upper Hubbard contribution at energy similar to the undoped case also shows good agreement with the ED results. 

%\begin{figure*}[htbp]

\begin{figure}[htbp]
  \begin{center}
   \includegraphics[width=80mm]{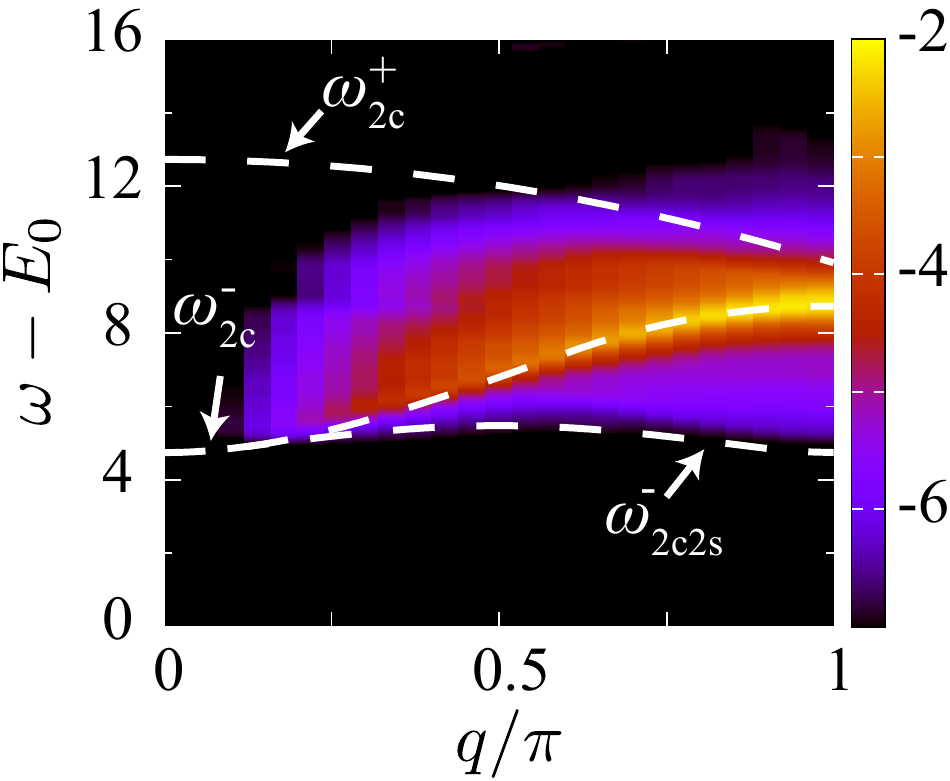}
  \end{center}
  \caption{(Color online) Contour plot of the charge dynamical structure factor $\log N(\bm{q},\omega)$ obtained by using the CF approach in the Hubbard chain for $L=50$ and $U/t=8$ at half filling.
  The result is plotted in the logarithmic scale in the same way as Fig. \ref{comparison}. 
  White dashed lines are dispersions characterizing the Bethe Ansatz 
solution as in Fig. \ref{comparison} (a). 
  }
  \label{chain_50}
\end{figure}

We also applied this method to a large size system 
that cannot be treated by the ED. The result is shown in Fig. \ref{chain_50}.
For large system size,
the broad continuum around $\omega-E_0 \sim U/t$ is clearly seen. 
We found that the strong lower edge and 
weak continuum below the edge for $q/\pi \geq 0.5$
appear in the spectrum, which are consistent 
with the previous DMRG calculation \cite{Pereira2012}.

%%%%%%%%%%%%%%%%%%%%%%%%%%%%%%%%%%%%%%%%%%%%%%%%%%%
%
\subsection{Two dimensional case}
\begin{figure}[htbp]
  \begin{center}
    \includegraphics[width=80mm]{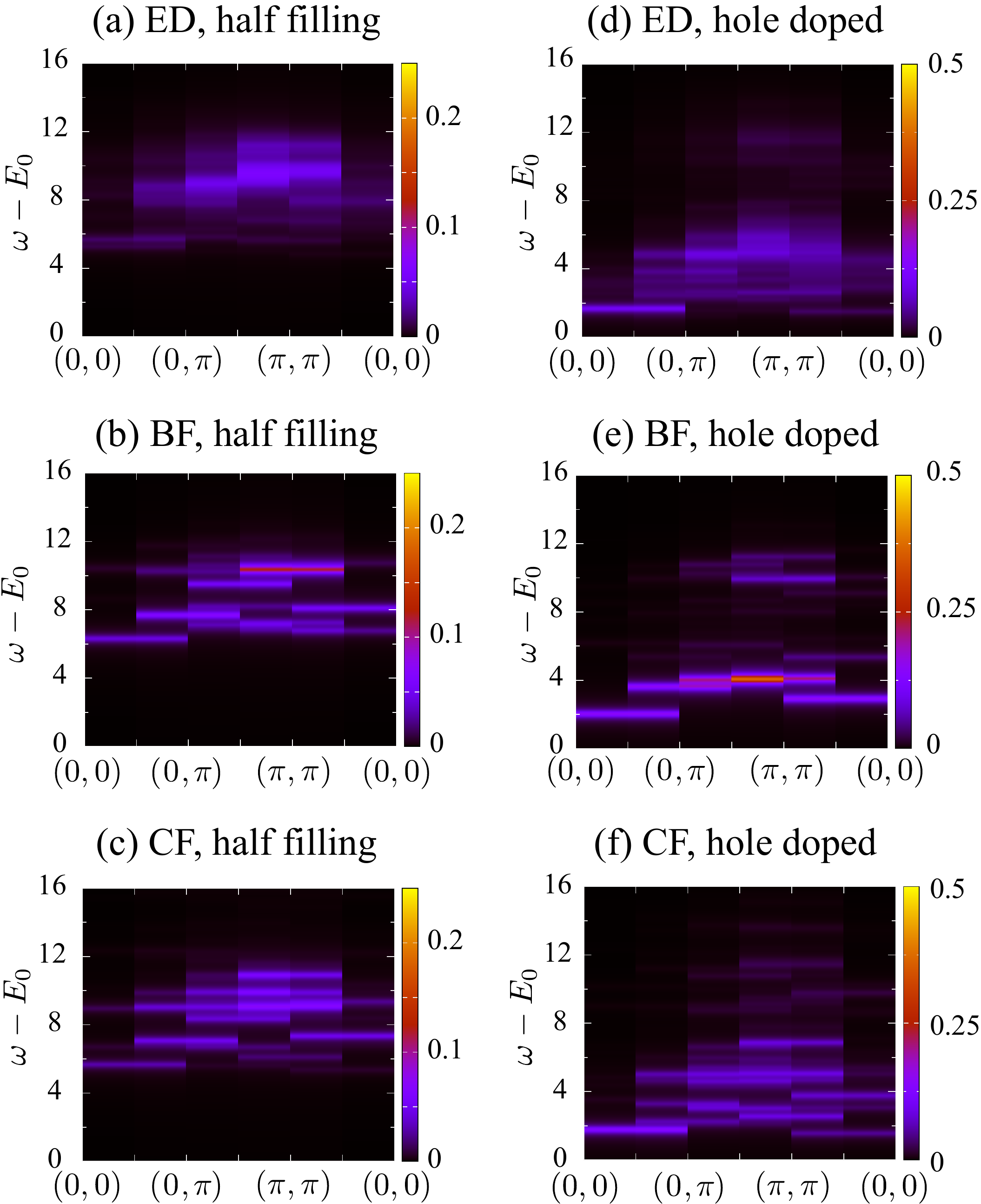}
  \end{center}
  \caption{(Color online) 
  Contour plot of the charge dynamical structure factor $N(\bm{q},\omega)$ in the two dimensional Hubbard model for $N_s=4 \times 4$ and $U/t=8$.
  $N(\bm{q},\omega)$ is measured along a high-symmetry path through the Brillouin zone.
  The results for $N_{\rm e}=16$ and $N_{\rm e}=12$ are shown in panels (a-c) and (d-f), respectively.
  Panels (a) and (d) are obtained by using the exact diagonalization method.
  Panels (b) and (e) show the results obtained by using the BF approach defined in Eq. (\ref{pair}). 
  Panels (c) and (f) show the results obtained by using the CF approach defined in Eq. (\ref{new}). 
  }
  \label{square}
\end{figure}

%
%%%%%%%%%%%%%%%%%%%%%%%%%%%%%%%%%%%%%%%%%%%%%%%%%%%%
In this subsection, we present the results for the Hubbard model 
on the square lattice for $U/t=8$.  In Fig.~\ref{square}, we show $N(\bm{q},\omega)$ on the high-symmetry line
for the system size $N_s=4 \times 4$. 
\ki{We} discuss the data at a larger system size \ki{later} in comparison to the particle-hole excitation speculated from the single-particle spectral function $A(\bm{k},\omega)$.
The boundary condition we used is the antiperiodic-periodic boundary 
condition to satisfy the closed shell condition at half filling. 

We first examine the accuracy of 
our variational descriptions at half filling as
shown in Figs.~\ref{square}(a)-(c).
At half filling, the ED result shows that
the strong but broad peak appears around $\bm{q}=(\pi,\pi)$ due to
the nearest-neighbor doublon-holon bindings. 
As is the case in one dimension, the BF 
approach does not describe charge dynamics even at half filling, 
i.e., the strong intensity at $\omega-E_0=8-10$ around $\bm{q}=(\pi,\pi)$ 
surrounded by broad structure is not well reproduced.
The CF approach clearly improves 
the description of this feature seen in the ED.

For hole-doped case in two dimensions, as shown in Figs.~\ref{square}(d)-(f), 
we again found that the CF approach
largely improves the charge dynamics compared with the BF approach.
Especially, the broad continuum \tfix{appears in} the CF approach in contrast to the BF approach.
These results show that the CF approach works well for describing the 
charge dynamics even in two dimensions both at half filling and the doped case.

\begin{figure}[htbp]
  \begin{center}
   \includegraphics[width=80mm]{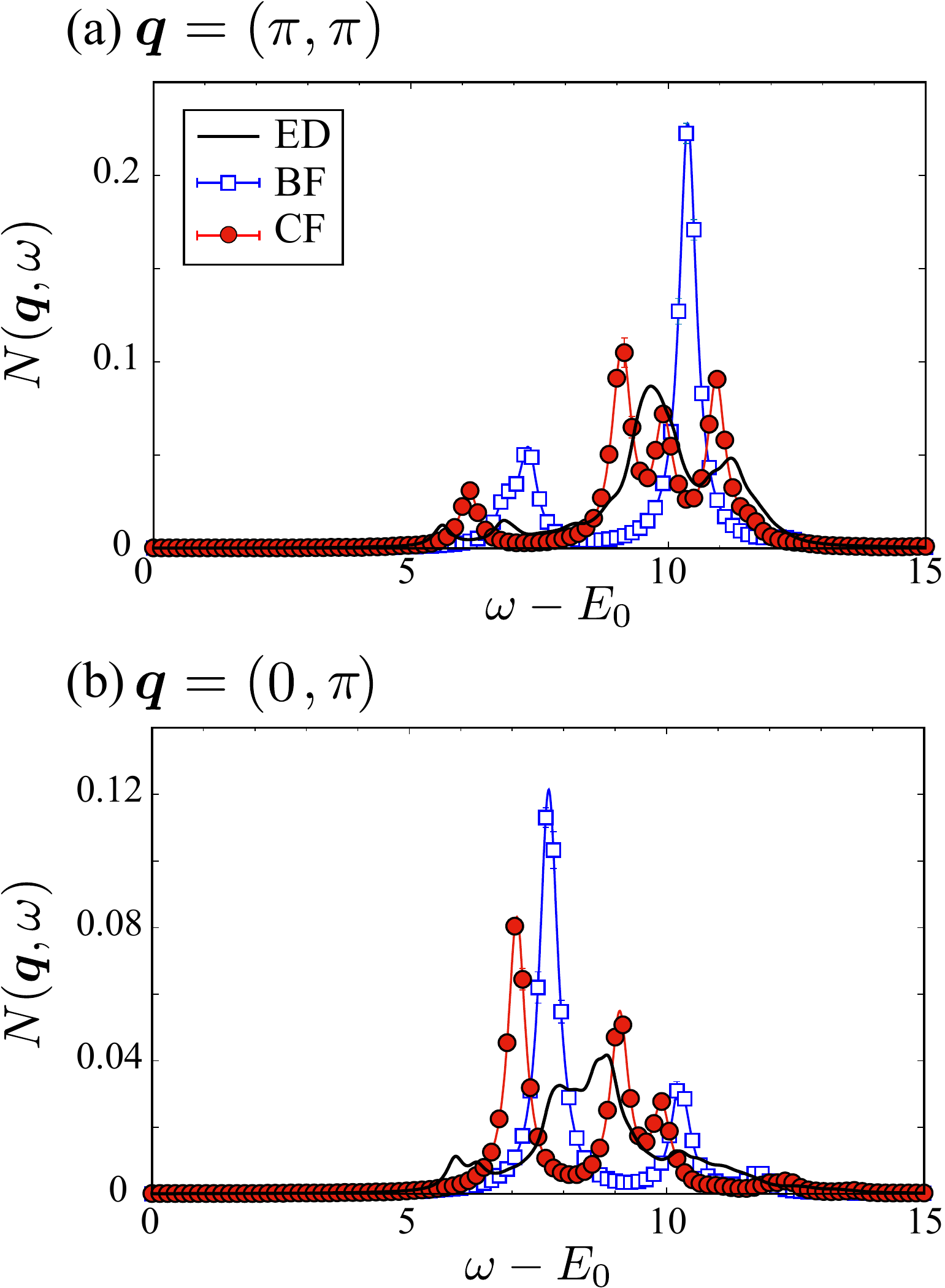}
  \end{center}
  \caption{
    (Color online) Charge dynamical structure factor for (a) $\bm{q}=(\pi,\pi)$ and (b) $\bm{q}=(0,\pi)$ in the two-dimensional Hubbard model for $L=4$ and $U/t=8$ at half filling. 
  Black solid lines are the results obtained by using the ED method.
  Open squares and solid circles represent the VMC results obtained by using the basis set in Eq. (\ref{pair}) and Eq. (\ref{new}), respectively. 
  }
  \label{qdep_square}
\end{figure}

There is, however, still discrepancies from the exact results. 
A discrepancy is clearly seen in Figs. \ref{qdep_square} (a) and (b), 
which shows $\omega$-dependence of $N(\bm{q},\omega)$ for $\bm{q}=(\pi, \pi)$ and $\bm{q}=(0, \pi)$, respectively. 
For $\bm{q}=(\pi, \pi)$, although only the single sharp peak appears around $\omega -E_0 \sim 10$ in the BF approach, 
this becomes broad structure in the CF approach and thus we can see the improvement by introducing the composite fermions. 
This improvement suggests that this broad structure in the ED result originates from the hybridization between the composite fermions defined in Eqs. (\ref{z_p}) and (\ref{z_m}).
For $\bm{q}=(0, \pi)$, however, 
the improvement by introducing the composite fermions is insufficient: 
The broad spectrum around $\omega -E_0 \sim 8$ in the ED result is not captured in both of the VMC results. 
This suggests existence of other $hidden$ fermions, which are required to take into account to capture charge dynamics in the two dimensional Mott insulator
as we discuss later.

\begin{figure}[htbp]
  \begin{center}
    \includegraphics[width=80mm]{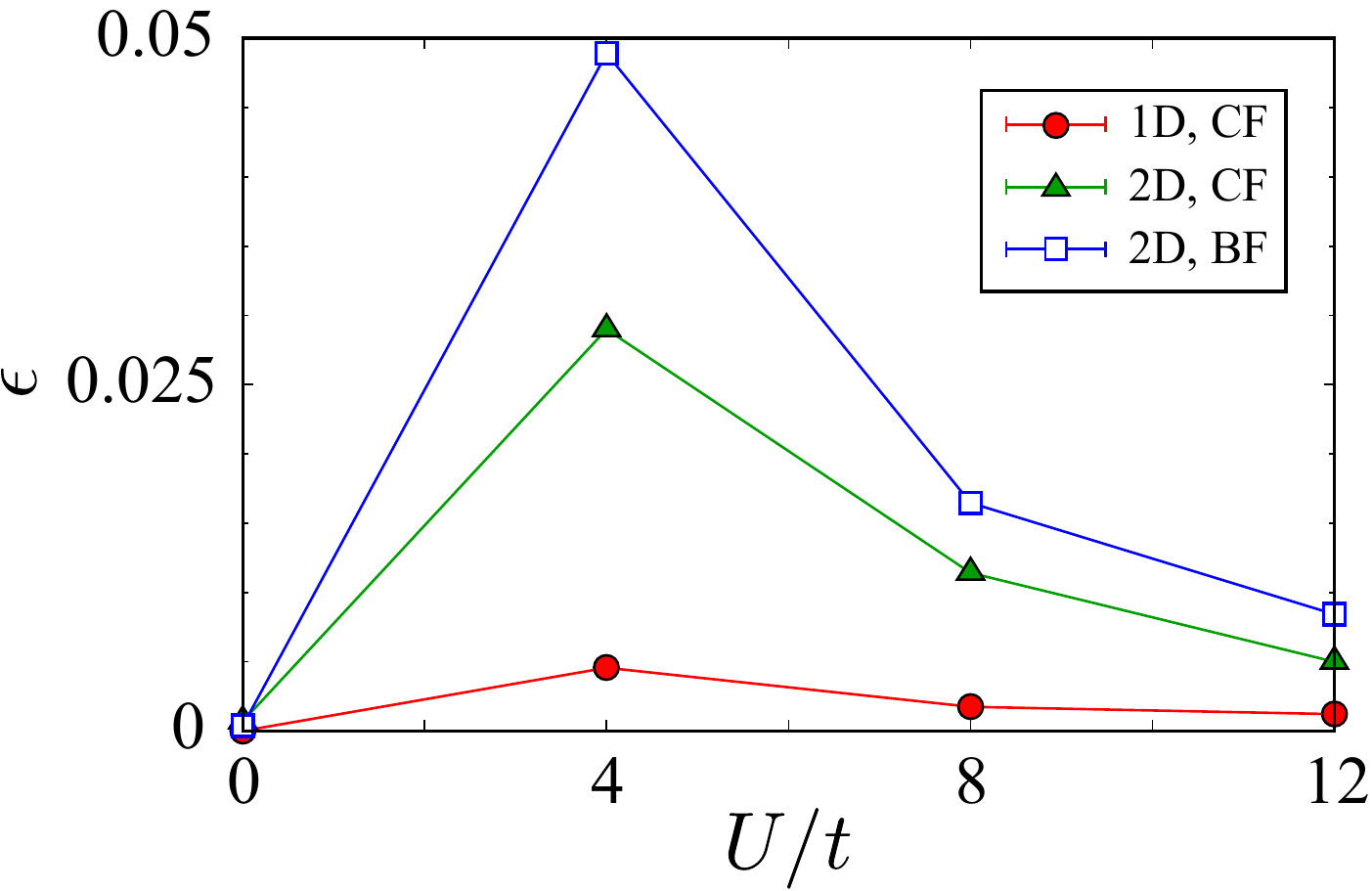}
  \end{center}
  \caption{
    (Color online) Interaction dependence of the mean absolute error $\epsilon_{\rm MAE}$ in the Hubbard model at half filling.
    The dimension of the system and the VMC approach we used are shown in the legend.
  }
  \label{diff_square}
\end{figure}

This problem is universally seen irrespective of the strength of the Coulomb interaction $U/t$.
To see that,
we plot the interaction dependence of the mean absolute error in Fig. \ref{diff_square}, which is defined by
\begin{align}
  \epsilon_{\rm MAE} = \frac{1}{N_{\bm{q}}} \sum_{\bm{q}}^{N_{\bm{q}}} \epsilon(\bm{q}),
\end{align}
where $N_{\bm{q}}$ is the number of $\bm{q}$ at which we measured $N(\bm{q},\omega)$.
We see that the introduction of the composite fermions 
reduces $\epsilon_{\rm MAE}$ for any strength of $U/t$, 
However, $\epsilon_{\rm MAE}$ for the two dimensional case obtained 
by the CF approach is still substantially larger than that for the one-dimensional system.
Our result indicates that the CF excitations 
may not be enough to describe the charge dynamics 
in the Mott insulators in higher dimensions. 
Identifying the key excitations to 
quantitatively describe the charge dynamics in 
the higher-dimensional Mott insulators 
is an intriguing issue, which will be discussed in the next section in detail.

\begin{figure}[htbp]
  \begin{center}
   \includegraphics[width=80mm]{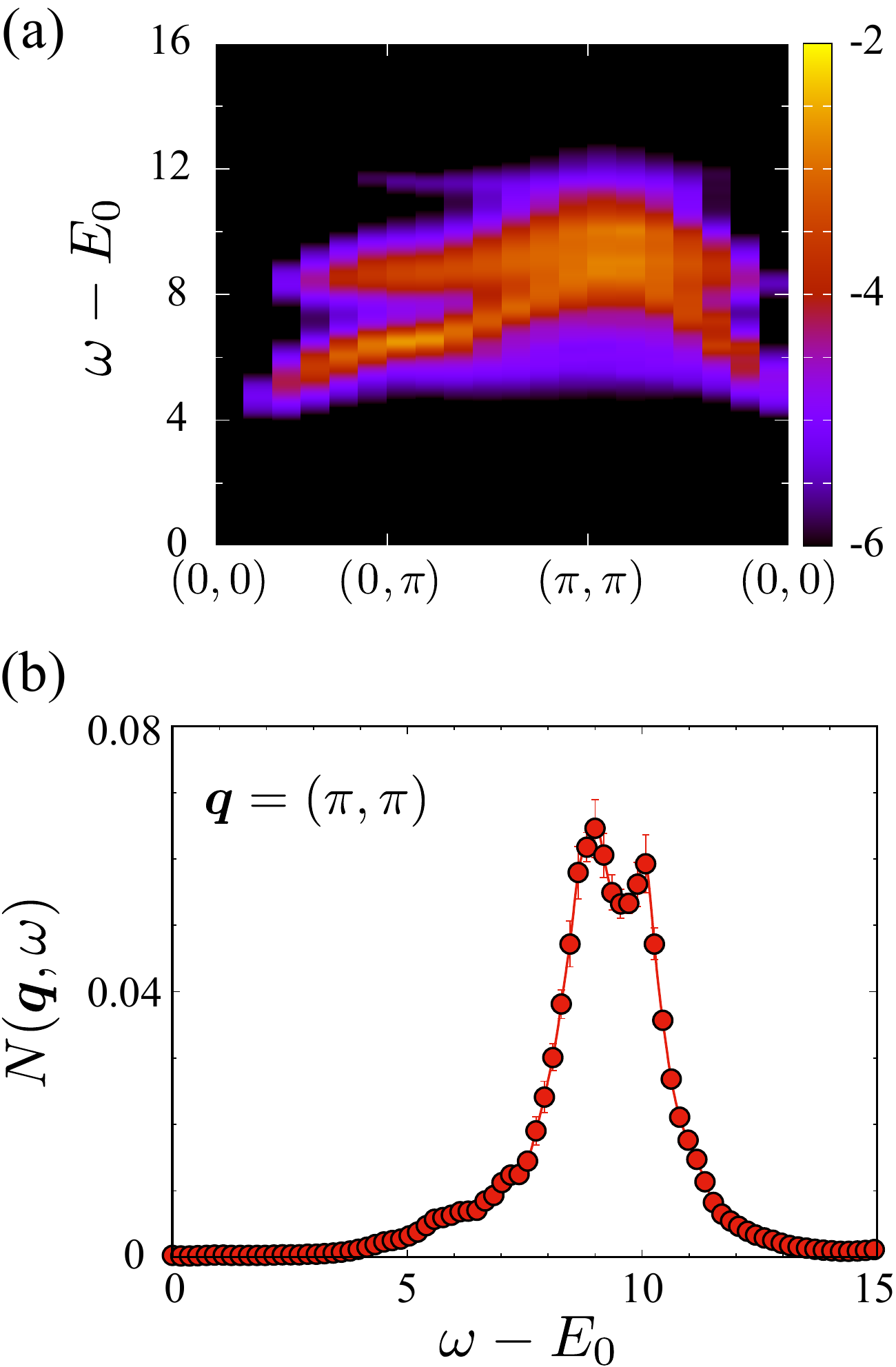}
  \end{center}
  \caption{
    (Color online) Charge dynamical structure factor obtained by VMC in the two-dimensional Hubbard model for $L=14$ and $U/t=8$ at half filling. 
    (a) Contour plot of $\log N(\bm{q},\omega)$.
    (b) $N(\bm{q},\omega)$ for $\bm{q}=(\pi,\pi)$.
  }
  \label{nq_square_large}
\end{figure}

\tfix{
  Nevertheless, the present results look correctly capture the particle-hole excitations from the lower Hubbard band to the upper Hubbard band inferred from 
  the single-particle spectral function $A(\bm{k},\omega)$ in Ref. \cite{Kohno2012,Charlebois2019}. 
  Particularly the highest intensity at $(\pi,\pi)$ with $\omega-E_0\sim 10-12$ in Fig. \ref{qdep_square}(a) is consistent with the particle-hole excitation from high-intensity lower Hubbard band 
  near $\Gamma$ to upper Hubbard band around $(\pi,\pi)$ points in $A(\bm{k},\omega)$ (see also Fig. \ref{nq_square_large} \ki{discussed below})}. 

\begin{figure}[htbp]
  \begin{center}
   \includegraphics[width=80mm]{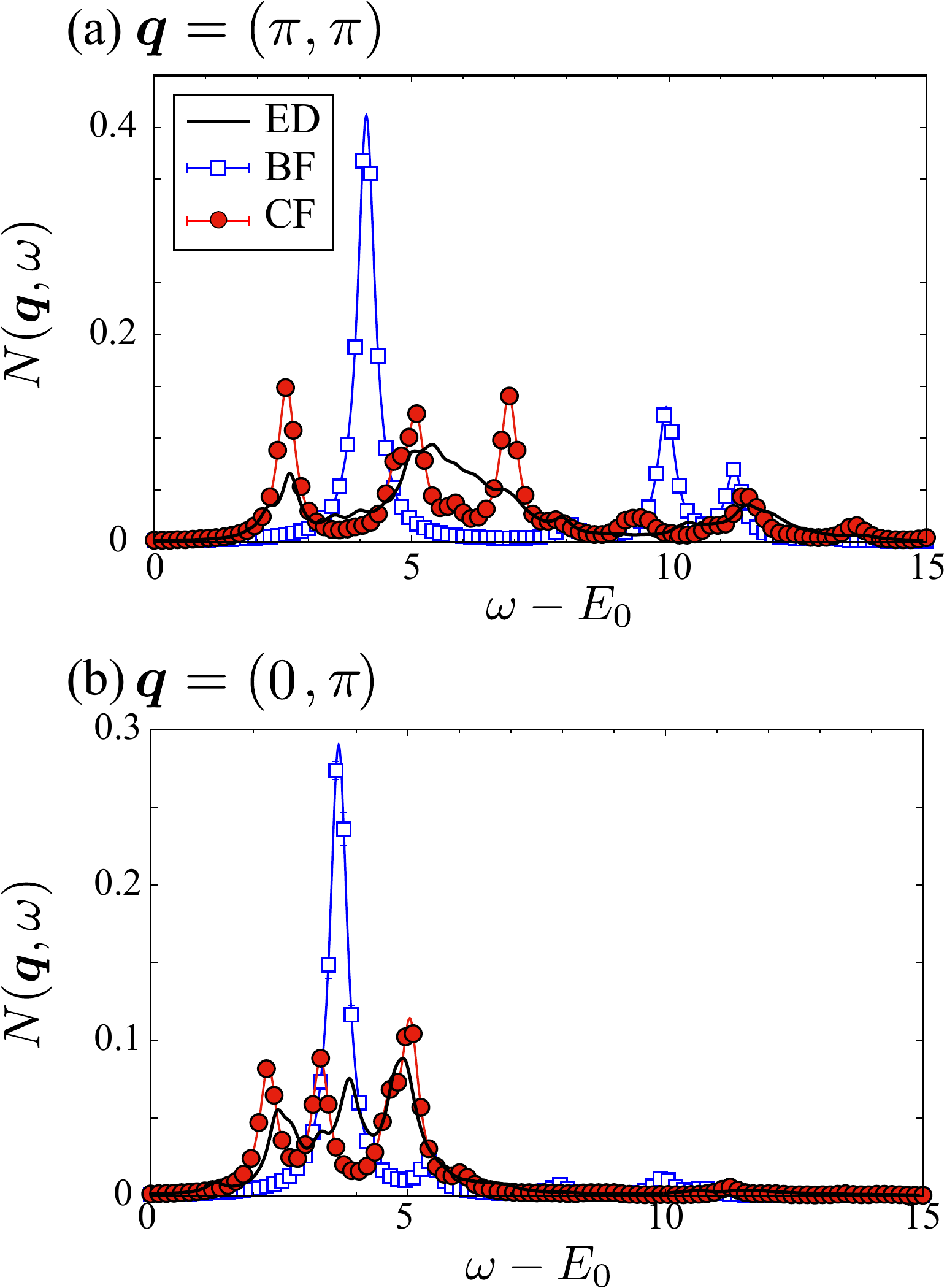}
  \end{center}
  \caption{
    (Color online) Charge dynamical structure factor for (a) $\bm{q}=(\pi,\pi)$ and (b) $\bm{q}=(0,\pi)$ in the two-dimensional Hubbard model for $L=4$, $U/t=8$ and $N_{\rm e}=12$. 
  Black solid lines are the results obtained by using the ED method.
  Open squares and solid circles represent the VMC results obtained by using the basis set in Eq. (\ref{pair}) and Eq. (\ref{new}), respectively. 
  }
  \label{qdep_square_hole}
\end{figure}

For the hole-doped case, we see that the CF approach reproduces the
overall charge dynamics. However, we notice that the intensities of $N(\bm{q},\omega)$ 
by the CF approach are also stronger than those by the ED, which is also found in Fig. \ref{qdep_square_hole} where $N(\bm{q},\omega)$ for $\bm{q}=(\pi,\pi)$ and $(0,\pi)$ is plotted.
The discrepancy for the incoherent part around $\omega-E_0 \sim 10$, especially for $\bm{q}=(\pi,\pi)$, may share the same origin for the insufficient description
of the charge dynamics in the Mott insulator at half filling as we have already shown. 
For $\omega-E_0 < 8$, the spectral weight shows a low-energy nearly flat dispersion in Fig.~\ref{square}(f). 
On the other hand the exact result in Fig.~\ref{square}(d) has much broader and damped feature. 
See also Fig. \ref{qdep_square_hole}, which shows that the broadness around $\omega-E_0 = 6$ for $\bm{q}=(\pi,\pi)$ and $\omega-E_0 = 4$ for $\bm{q}=(0,\pi)$ is underestimated 
and the intensity for $\omega-E_0 \sim 2$ and $\bm{q}=(\pi,\pi)$ is overestimated.
We discuss the possible origin of the discrepancy in the next section.

\ki{Finally, we show $N(\bm{q},\omega)$ in the two-dimensional Hubbard model for $L=14$ and $U/t=8$ at half filling in Fig. \ref{nq_square_large} as reference data for future studies. 
We see the broad structure with strong intensity around $\omega-E_0 \sim 10$ in Fig. \ref{nq_square_large}(b), which is consistent with the particle-hole excitation inferred from $A(\bm{k},\omega)$ 
\ki{as we have discussed above in connection to Fig. \ref{qdep_square}(a)}.
We also find that, below $\omega - E_0 \sim 8$ around $(\pi, \pi)$, there is broad continuum, which is similar to the case for the two-holon-two-spinon [or two-doublon-two-spinon] in the one-dimensional Hubbard model.
Further investigations of the similarity/difference between the charge dynamics in one- and two-dimensional systems are an important issue, but left for future studies.
}

%%%%%%%%%%%%%%%%%%%%%%%%%%%%%%%%%%%%%%%%%%%%%%%%%%%
%
\section{Discussions and summary}\label{sec:summary}
%
%%%%%%%%%%%%%%%%%%%%%%%%%%%%%%%%%%%%%%%%%%%%%%%%%%%%

The proposed VMC approach incorporating the composite fermion excitations provides us with
accurate charge dynamical structure factors in one dimension and correctly reproduces signature of spin-charge separation with broad incoherent continuum.
It also shows a \tfix{qualitatively good agreement} %fairly good agreement 
with the ED results in two dimensions, 
with broad continuum and large intensity around $(\pi,\pi)$ in the incoherent part. 
However, we still have rooms for improvement in both the undoped case and the doped Mott insulator in two dimensions.

Here, we discuss the possible ways 
to improve the charge dynamics in two dimensions.
There are two possible reasons for the insufficient 
broad continuum in the two dimensional system. 
One possible origin is the insufficient accuracy of the ground state,
 because the relative error of the ground-state energy tends to be 
larger than that in the chain~\cite{Tahara2008a,Kaneko2013}.
Recent proposed complementary methods for quantum lattice models 
such as the introductions of backflow correlations~\cite{Tocchio2008,Tocchio2011,Ido2015} 
and tensor network~\cite{Chou2012,Sikora2015,Zhao2017} could 
improve the Jastrow-type trial wavefunctions.

Second is the limitation of the assumed restricted basis set for the excitation 
in extracting the correct broad continuum. 
To enlarge the restricted subspace, we would need to introduce an efficient basis set for excited states in two dimensional systems. 
\tfix{An important process missing in our approach is scattering between composite fermions through many-body interactions. 
This effect would be taken into account by adding multiple particle-hole excitations in the basis set.
}
A simple\tfix{r} candidate \tfix{than inclusion of multiple particle-hole excitations} is another composite fermion which 
depends on intersite configurations.
Such a composite fermion is required to consider charge dynamics 
in the antiferromagnetic background with spin fluctuations in the two-dimensional systems, suggested in previous studies\cite{Avella2003,Odashima2005}. 

The present construction of excited states is able to describe the upper and lower Hubbard excitations correctly because of the form Eqs.(\ref{z_p}) and (\ref{z_m}). 
However, in the doped system, ingap states are known to emerge, which also generates the pseudogap structure~\cite{Ohta1992,Leung1992,Dagotto1992,Preuss1995,Preuss1997,Kyung2006,Sakai2009}. 
Such low-energy excitations near the Fermi level are not explicitly contained in the construction 
Eqs. (\ref{z_p}) and (\ref{z_m}) and likely to fail in capturing this emergent structure in the present form.
It was proposed that such ingap states are generated by the coupling to excitonic excitations with weak binding energy in contrast to the doublon-holon binding generating the large Mott gap\cite{Imada2019}. 
Such weakly bound excitons are moreover hybridizing with the spin singlet with the resonating valence bond nature~\cite{Anderson1987} 
and may give distinct ingap structure. 
This idea offers an alternative view to the coupling to the spin fluctuation in the antiferromagnetic background mentioned above.
Both views suggest the importance to take into account long-ranged entanglement or correlation (from the nearest neighbour to several-distance neighbors) of the composite fermion
such as 
$d_{j,\sigma}^{\dagger}= \sum_{\delta}\tilde{d}^{\dagger}_{j,\delta,\sigma}$
with variational parameters $\kappa_{\delta}, g_{\delta}, \alpha_{\delta}, \beta_{\delta}$ and $\gamma_{\delta}$, defined as
\begin{eqnarray}
  \tilde{d}^{\dagger}_{j,\delta,\sigma}&=& K_{j,\delta,\sigma}c^{\dagger}_{j,\overline{\sigma}} + G_{j,\delta,\sigma}c^{\dagger}_{j,\sigma}, \\
  K_{j,\delta,\sigma}&=&\kappa_{\delta}c^{\dagger}_{j+\delta,\sigma}c_{j+\delta,\overline{\sigma}}, \\
  G_{j,\delta,\sigma}&=&g_{\delta}-\alpha_{\delta}n_{j+\delta,\overline{\sigma}}-\beta_{\delta}n_{j+\delta,\sigma}+\gamma_{\delta}n_{j+\delta,\sigma}n_{j+\delta,\overline{\sigma}}, \nonumber \\ 
\end{eqnarray}
involving neighboring sites $j+\delta$ to $j$ as suggested in the context of the dark fermion in Ref.\cite{Imada2019}.

Both possibilities of improving the ground and excited states are desired to be examined toward more quantitative 
understanding of intriguing phenomena in 
quantum many-body systems. It is a fundamentally important issue for future studies.

We also note that the present approach would be also useful to obtain other 
dynamical physical quantities such as the optical 
conductivity $\sigma(\omega)$ and 
the one-particle spectral function $A(\bm{k},\omega)$.
These extensions will be studied in the near future.

\tfix{Another intriguing issue is to calculate $N(\bm{q},\omega)$ of the $d$-wave superconducting state in two dimensional correlated electron systems with large system sizes.
In RIXS measurements, it was reported that dynamical charge fluctuations pervaded the phase diagram of a cuprate superconductor\cite{Arpaia2019}, 
which implies that the charge fluctuations play an important role on the emergence of high-$T_{\bm c}$ superconductivity.
If the strength of the dynamical charge fluctuations directly determines $T_{\bm c}$, 
inclusions of particle-particle and hole-hole excitations in the basis set would be important when we analyze $N(\bm{q},\omega)$. 
This will be also examined in the future study.}

In summary, 
we examined how the VMC approach based on the Li-Yang method
can be extended to describe the charge dynamics in the 
one- and two-dimensional Hubbard models.
We found that the CF excitations are important for describing 
the charge dynamics in the Hubbard model.
In the one-dimensional Hubbard model,
we have shown that the CF approach well
reproduces the results by the exact diagonalization
at half filling and the hole-doped case as well.
In the two-dimensional Hubbard model,
although the CF approach largely improves the BF results, 
quantitative discrepancy from the exact results still exists.
Our results indicate that the excitations complementing the present local CF 
excitations are necessary for quantitative description of the charge dynamics
in the two-dimensional Mott insulator.
This is consistent with the emergence of the ingap states and the pseudogap,
which are ignored in the present consideration of the local excitation.
Spatially more extended and spin dependent composite fermions such as the dark fermion (or hidden fermion)~\cite{Sakai2016,Sakai2018,Imada2019} 
must be involved in the Hilbert space for the excitation. 
The role of coupling to weakly bound exciton is an intriguing future issue.

It is also an intriguing future study to 
apply our method to {\it ab initio} Hamiltonians
for high-$T_{\rm c}$ superconductors~\cite{Miyake2010,Hirayama2018,Nilsson2019}
and clarify how the enhanced dynamical 
charge fluctuations including the uniform 
static charge fluctuations
affects high-$T_{\rm c}$ superconductivity
~\cite{Misawa2014a,Misawa2014,Ohgoe2019}. 

%%%%%%%%%%%%%%%%%%%%%%%%%%%%%%%%%%%%%%%%%%%%%%%%%%%%
%
{\it Acknowledgments.}---
%
%%%%%%%%%%%%%%%%%%%%%%%%%%%%%%%%%%%%%%%%%%%%%%%%%%%%
Our VMC code has been developed based on open-source software ``mVMC"\cite{Misawa2017}.
To obtain the exact diagonalization results, 
we used ``$\mathcal{H} \Phi$" package\cite{Kawamura2017}. 
KI thanks Takahiro Ohgoe for continuous discussion. 
We acknowledge Maxime Charlebois and Youhei Yamaji for 
fruitful discussions and important suggestions.
We thank the Supercomputer Center, the Institute 
for Solid State Physics, the University of Tokyo for the facilities. 
The authors are grateful to the support by a Grant-in-Aid for Scientific Research 
(Nos. 16K17746, 16H06345, 19K03739, 19K14645)
from Ministry of Education, Culture, Sports, Science and Technology, Japan.  
TM and KI were supported by Building of Consortia for the Development of 
Human Resources in Science and Technology from the MEXT of Japan. 
This work is financially 
supported by the MEXT HPCI Strategic Programs, and the 
Creation of New Functional Devices and High-Performance 
Materials to Support Next Generation Industries (CDMSI). 
We also acknowledge the support 
provided by the RIKEN Advanced Institute for Computational
Science under the HPCI System Research project 
(Grants No. hp170263, hp180170 and hp190145).
%%%%%%%%%%%%%%%%%%%%%%%%%%%%%%%%%%%%%%%%%%%%%%%%%%%
%%%%%%%%%%%%%%%%
%%%%%%%%%%%%%%%%%%%%%%%%%%%%%%%%%%%%%%%%%%%%%%%%%%%%
%
%              APPENDIX
%
%%%%%%%%%%%%%%%%%%%%%%%%%%%%%%%%%%%%%%%%%%%%%%%%%%%%
%\clearpage
\appendix
%\if0
%{\LARGE Supplemental  Materials for ``''}
 
%\renewcommand{\theequation}{S.\arabic{equation}}
%\setcounter{equation}{0}
%\renewcommand{\theequation}{S.\arabic{equation}}
%\renewcommand{\tablename}{Table S}
%%%%%%%%%%%%%%%%%%%%%%%%%%%%%%%%%%%%%%%%%%%%%%%%%%%%
%
\section{Reweighting technique}\label{appendixA}
%
%%%%%%%%%%%%%%%%%%%%%%%%%%%%%%%%%%%%%%%%%%%%%%%%%%%%
In this appendix, we introduce the reweighting technique to efficiently evaluate the matrix element of Hamiltonian and overlap matrices, Eqs. (\ref{rh}) and (\ref{ro}) respectively, as proposed by Li and Yang\cite{Li2010}.
By using the reweighting technique, we evaluate Eqs. (\ref{rh}) and (\ref{ro}) as
\begin{align}
  &O^{\bm{q}}_{nm}
  = \frac{\sum_x \braket{\bm{q},n | x} \braket{x| \bm{q},m}}{\sum_x \sum_n |\braket{\bm{q},n | x}|^2 } \left(\frac{\sum_x \braket{\psi | x} \braket{x| \psi}}{\sum_x \sum_n |\braket{\bm{q},n | x}|^2} \right)^{-1}\\
  %&=& \frac{\sum_x W(x) \braket{\psi_n | x} \braket{x| \psi_m} W^{-1}(x)}{\sum_x W(x)}, \\
  & \approx %\frac{1}{N_{\rm smp}} \sum_i \frac{ \braket{\bm{q},n | x_i} \braket{x_i| \bm{q},m}} {W(x_i)} \\
  %&= 
  \frac{1}{N_{\rm smp}} \sum_i o^*_n(x_i)  o_m(x_i) \left( \frac{1}{N_{\rm smp}} \sum_i \frac{ |\braket{x| \psi}|^2 }{W(x)}\right)^{-1}, \\
  &o_n(x_i) = \frac{ \braket{x_i| \bm{q},n} }{ \sqrt{W(x_i)} }, \\ 
  &W(x)=\sum_n |\braket{\bm{q},n| x}|^2
  %%%%%
\end{align}
and
\begin{align}
  &H^{\bm{q}}_{nm}
  \approx %\frac{1}{2N_{\rm smp}} \sum_i \frac{ \braket{\psi_n | x_i} \braket{x_i| \mathcal{H} |\psi_m} + \braket{\psi_n | \mathcal{H} | x_i} \braket{x_i| \psi_m}} {W(x_i)} \\
  %&= 
  \frac{1}{2N_{\rm smp}} \sum_i \left( o^*_n(x_i)  h_m(x_i) +  h^*_n(x_i)  o_m(x_i)\right) \nonumber \\ 
  & \ \ \ \  \cdot \left( \frac{1}{N_{\rm smp}} \sum_i \frac{ |\braket{x| \psi}|^2 }{W(x)}\right)^{-1}, \\
  &h_n(x_i) = \frac{ \braket{x_i| \mathcal{H} | \bm{q},n} }{ \sqrt{W(x_i)} }, \label{ham}
  %%%%%
\end{align}
respectively. Therefore, we calculate up to sixth-order correlation functions to evaluate $h_n(x_i)$ and $o_n(x_i)$ at each sample.
Since the weight $W(x)$ is dependent on all bases we considered, we can reduce the statistical error caused by the node-difference among $\braket{x | \bm{q},n}$.
See also Ref. \cite{Li2010}, where the reweighting technique is discussed in details.

%%%%%%%%%%%%%%%%%%%%%%%%%%%%%%%%%%%%%%%%%%%%%%%%%%%%
%
\section{Eta-dependence of $N(\bm{q},\omega)$}\label{appendixB}
%
%%%%%%%%%%%%%%%%%%%%%%%%%%%%%%%%%%%%%%%%%%%%%%%%%%%%
In this appendix, we show the eta-dependence of $N(\bm{q},\omega)$. 
In this paper, we set the smearing factor $\eta=0.2$. 
To capture some of the features of the spectrum, It is better to take $\eta$ smaller than typical energy scales such as effective exchange interaction $J$ induced by the strong onsite interaction.
This condition is barely satisfied for $U/t=8$ because the effective exchange interaction for $U/t=8$ is $J/t =4t/U=0.5$.

\begin{figure}[htbp]
  \begin{center}
   \includegraphics[width=60mm]{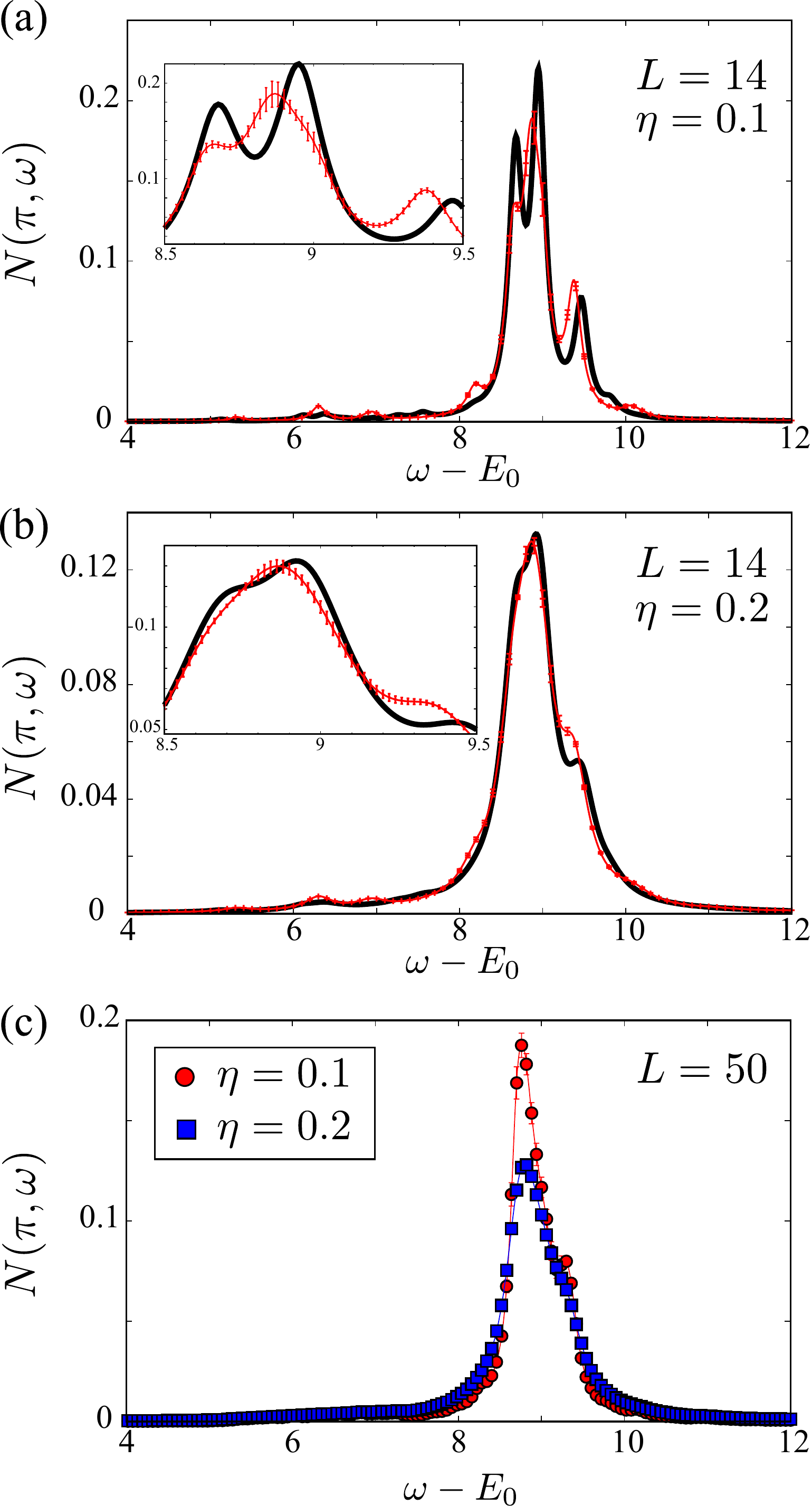}
  \end{center}
  \caption{
    (Color online) $\eta$-dependence of $N(\pi,\omega)$ in one dimensional system with $U/t=8$.
    (a) $N(\pi,\omega)$ for $L=14$ and $\eta=0.1$.
    Black thick and red thin lines represent ED and VMC results, respectively.
    Insets are the enlarged spectrum around $\omega-E_0 = 9$.
    (b) $N(\pi,\omega)$ for $L=14$ and $\eta=0.2$. Notations are the same as in the panel (a).
    (c) $N(\pi,\omega)$ for $L=50$. Red circles and blue squares are the VMC results for $\eta=0.1$ and 0.2, respectively. 
  }
  \label{etadep_app}
\end{figure}

To confirm whether our employed $\eta$ is appropriate, we have calculated $N(\bm{q},\omega)$ in one dimensional system with $U/t=8$ for $\eta=0.1$, $0.2$ and $0.5$ 
which are shown in Fig. \ref{etadep_app}. The spectrum around $\omega-E_0 = 9$, which is the major part of the spectrum, is plotted as the insets for each panel.
For $\eta=0.1$, VMC results reproduce two clear peaks around $\omega-E_0=8.7$ and $9.4$ in the exact results.
This prominent feature is still found as the existence of shoulders in the same $\omega$ range for $\eta=0.2$ in Fig. \ref{etadep_app}(b) and even in larger size systems as shown in Fig. \ref{etadep_app}(c).
This result indicates that $\eta$ = 0.2 result captures the major features in the charge dynamical structure factors.
In fact the shoulder at $\omega-E_0\sim 8.7$ with the shoulder at 9.4 is consistent with the spinon-holon band splitting originated from the spin-charge separation 
(see Refs. \cite{Kim1996, Kim1997}) which should emerge as the splitting in the particle-hole excitation here at momentum $\pi$.

%%%%%%%%%%%%%%%%%%%%%%%%%%%%%%%%%%%%%%%%%%%%%%%%%%%%
%
%\section{$N(\bm{q},\omega)$ for other wavenumbers}\label{appendixC}
\section{Supplemental results for $N(\bm{q},\omega)$}\label{appendixC}
\begin{figure}[htbp]
  \begin{center}
   \includegraphics[width=80mm]{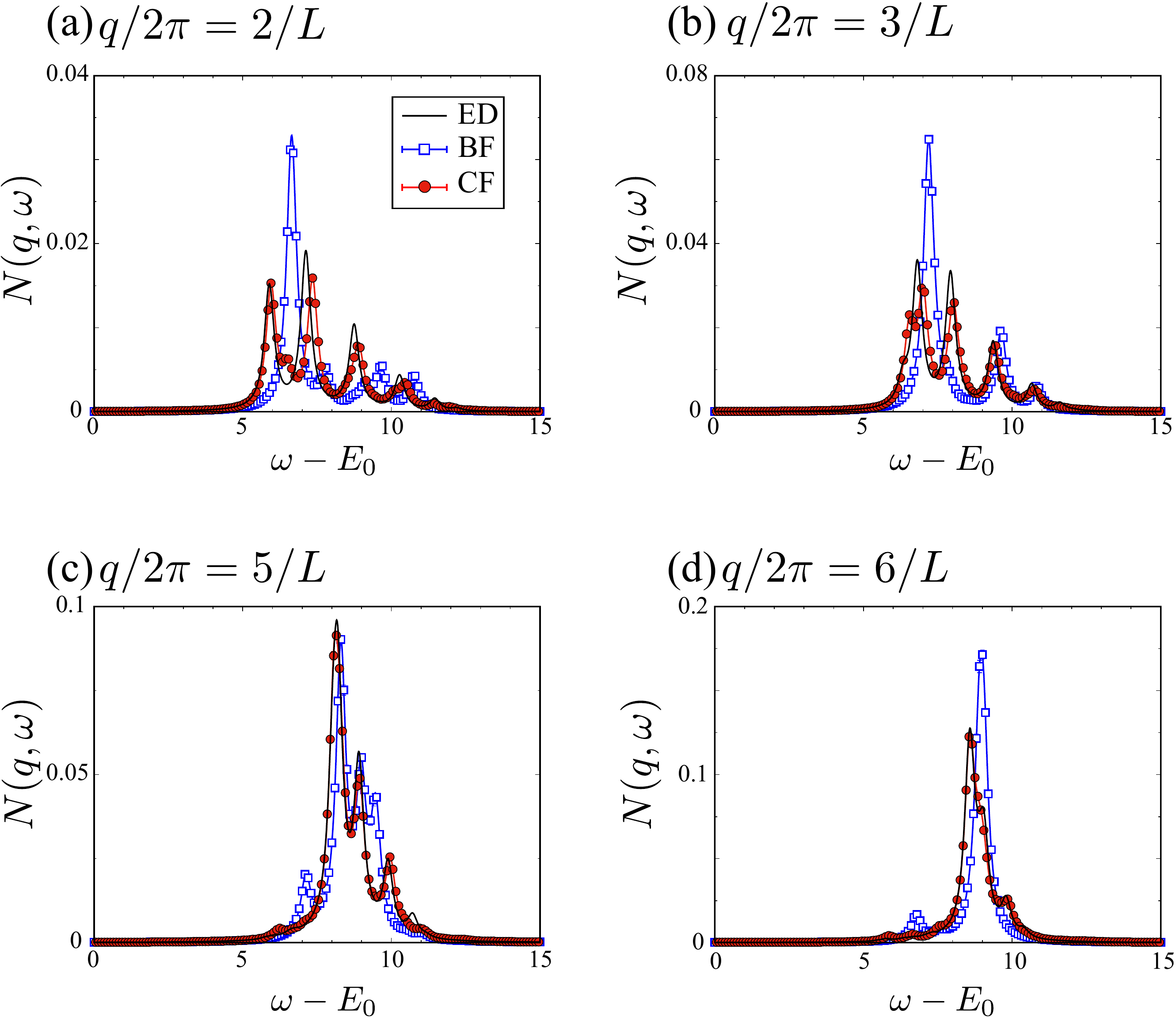}
  \end{center}
  \caption{
    (Color online) Charge dynamical structure factor for several $q$ in the one-dimensional Hubbard model for $L=14$ and $U/t=8$ at half filling. 
    Notations are the same as in Fig. \ref{qdep_chain}. 
  }
  \label{qdep_chain_app}
\end{figure}

\begin{figure}[htbp]
  \begin{center}
   \includegraphics[width=60mm]{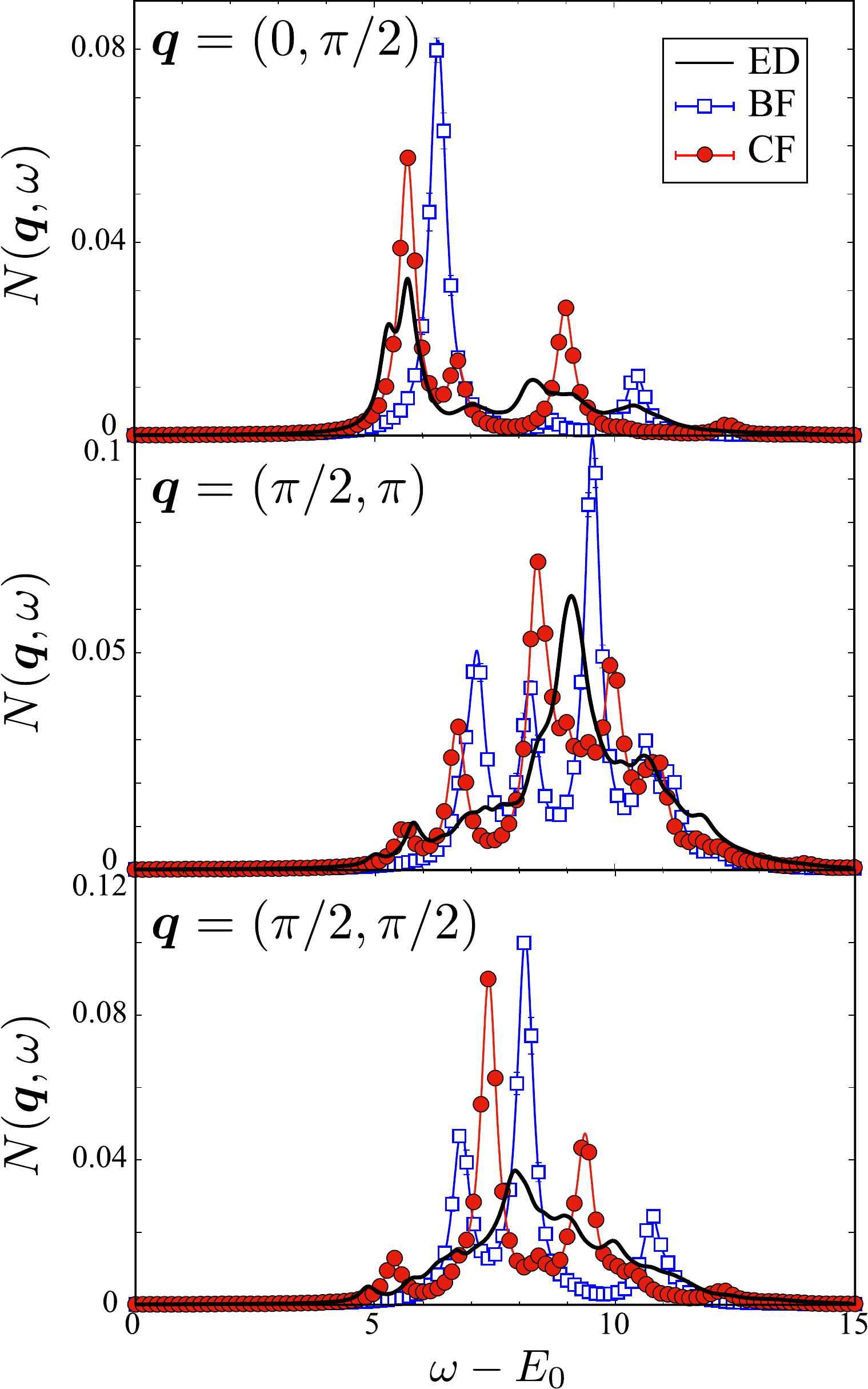}
  \end{center}
  \caption{
    (Color online) Charge dynamical structure factor for $\bm{q}=(0, \pi/2)$, $\bm{q}=(\pi/2,\pi)$ and $\bm{q}=(\pi/2,\pi/2)$ in the two-dimensional Hubbard model for $L=4$ and $U/t=8$ at half filling. 
    Notations are the same as in Fig. \ref{qdep_square}. 
  }
  \label{qdep_square_app}
\end{figure}

\begin{figure}[htbp]
  \begin{center}
   \includegraphics[width=60mm]{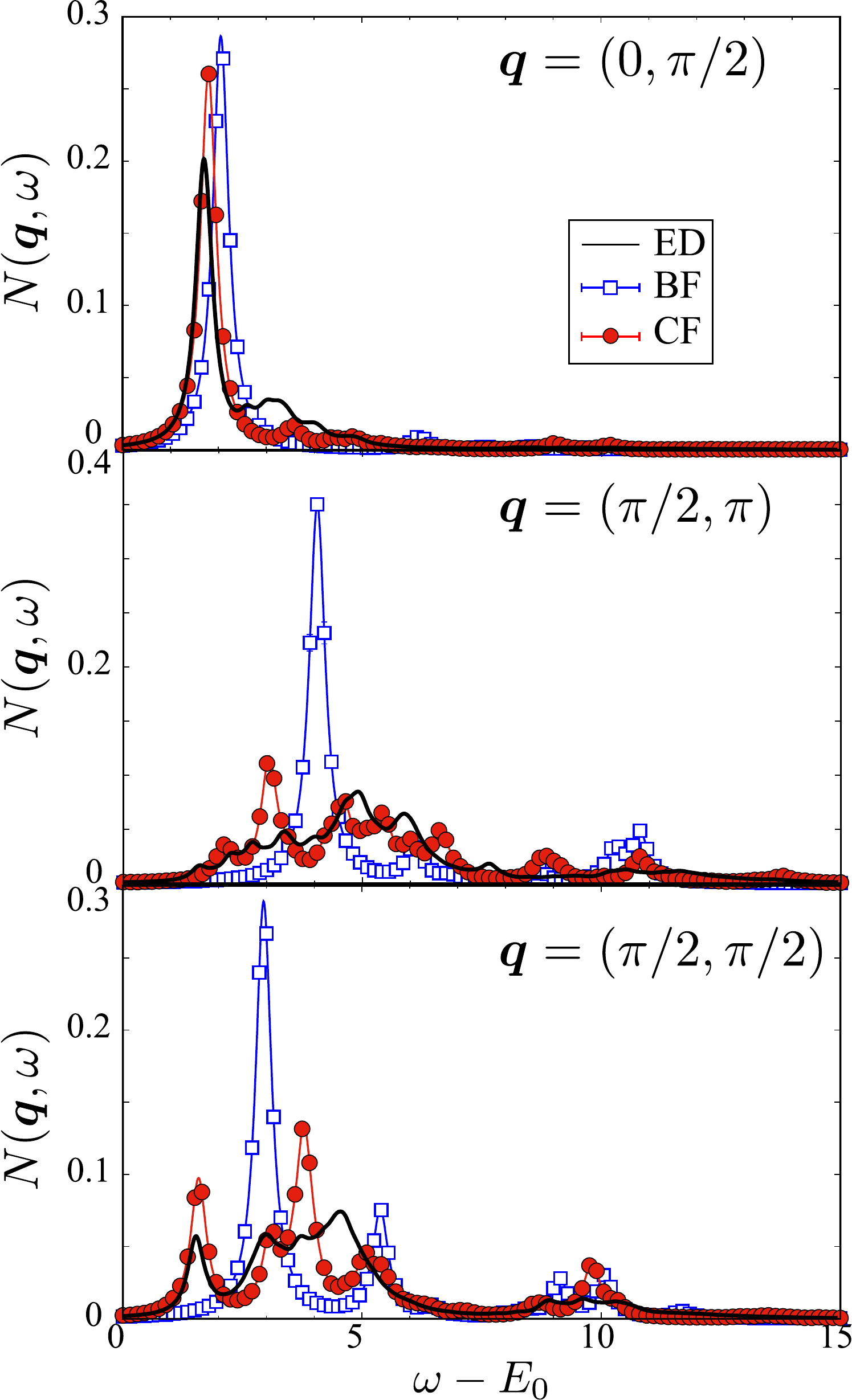}
  \end{center}
  \caption{
    (Color online) Charge dynamical structure factor for $\bm{q}=(0,\pi/2)$, $\bm{q}=(\pi/2,\pi)$ and $\bm{q}=(\pi/2,\pi/2)$ in the two-dimensional Hubbard model for $L=4$, $U/t=8$ and $N=12$. 
    Notations are the same as in Fig. \ref{qdep_square_hole}. 
  }
  \label{qdep_square_hole_app}
\end{figure}

%
%%%%%%%%%%%%%%%%%%%%%%%%%%%%%%%%%%%%%%%%%%%%%%%%%%%%
As supplemental results, we plot Figs. \ref{qdep_chain_app}, \ref{qdep_square_app} and \ref{qdep_square_hole_app}, which show $N(\bm{q},\omega)$ in one- and two-dimensional Hubbard model 
at other wavenumbers that are not shown in Figs. \ref{qdep_chain}, \ref{qdep_square} and \ref{qdep_square_hole}, respectively.
Aside from detailed and precise energies of the peaks, the peak structures and the weight show good agreements.

\clearpage

%\fi
%---------------------------------------------------
\bibliographystyle{prsty}
\bibliography{reference}
%---- Backmatter ----
%\backmatter
%\nocite{*}
%\bibliographystyle{unsrt}
\end{document}